\documentclass[11pt]{article}

\usepackage{latexsym}
\usepackage{amsmath}
\usepackage{amssymb}
\usepackage{graphicx}
\usepackage{mathrsfs}
\usepackage{bm}
\usepackage{amstext,amssymb,amsmath,float}
\usepackage{epsfig,subfigure}
\usepackage{yfonts}

\makeatletter

\small\normalsize
\advance \textheight by \topskip

\ifcase \@ptsize
\or
\or
\fi

\makeatother

\newenvironment{arrayl}%
        {\renewcommand{\arraystretch}{1.5}
        \begin{array}{@{}r@{\hskip\arraycolsep}c@{\hskip\arraycolsep}l}}%
        {\end{array}\renewcommand{\arraystretch}{1}}
\newtheorem{lemma}{Lemma}
\newtheorem{theorem}[lemma]{Theorem}

\newtheorem{definition}[lemma]{Definition}

\newtheorem{conjecture}{Conjecture}
\newtheorem{remar}{Remark}
\newtheorem{examp}{Example}

\numberwithin{equation}{section}

\newcommand{\Be}{\mathbf{e}}
\newcommand{\bM}{\mathbf{M}}
\newcommand{\bA}{\mathbf{A}}
\newcommand{\ba}{\mathbf{a}}
\newcommand{\bP}{\mathbf{P}}
\newcommand{\bI}{\mathbf{I}}
\newcommand{\bU}{\mathbf{U}}
\newcommand{\bu}{\mathbf{u}}
\newcommand{\bw}{\mathbf{w}}
\newcommand{\bW}{\mathbf{W}}

\newcommand{\bPhi}{\mathbf{\Phi}}
\newcommand \be{\begin{eqnarray}}
\newcommand \ee{\end{eqnarray}}
\newcommand{\qmbox}[1]{\quad\mbox{#1}\quad}
\newcommand{\real}{{\mathbb{R}}}

\def\mijnlim#1{\mathop{\rm lim}\limits_{#1}}
\def\eps{\varepsilon}
\def\diag{\mathop{\rm diag}}
\def\rrbf#1{{#1}} 
\def\rbf#1{{#1}} 
\def\rrul#1{{#1}} 
\def\rul#1{{#1}} 

\title{\bf Laplacian instability of planar streamer ionization fronts ---
   an example of pulled front analysis\thanks{G. Derks acknowledges a travel
    grant of the Royal Society, which initiated this research, and a
    visitor grant of the Dutch funding agency NWO and the NWO-mathematics
    cluster NDNS${}^+$ to finish the work. The work was also supported by a CWI
    PhD grant for B. Meulenbroek.}}

\author{Gianne Derks\thanks{Department of Mathematics,
University of Surrey, Guildford, Surrey, GU2 7XH, UK (G.Derks@surrey.ac.uk).} \and Ute Ebert\thanks{Cluster
`Modelling, Analysis and Simulation', Center for Mathematics and Computer Science (CWI), P.O. Box 94079, 1090
GB Amsterdam, and also at Eindhoven Univ.\ Techn., The Netherlands (ebert@cwi.nl).} \and Bernard
Meulenbroek\thanks{While contributing to this paper, at CWI Amsterdam, now at Faculty of Electrical
Engineering, Mathematics and Computer Science, Delft Univ.\ Techn., P.O. Box 5031, 2600 GA Delft, The
Netherlands (B.J.Meulenbroek@tudelft.nl).}}

\begin{document}

\maketitle

\begin{abstract}\noindent

  Streamer ionization fronts are pulled fronts propagating into a
  linearly unstable state; the spatial decay of the initial condition
  of a planar front selects dynamically one specific long time
  attractor out of a continuous family. A \rbf{stability analysis for
    perturbations in the transverse direction} has to take these
  features into account. In this paper we \rbf{show how to apply the
    Evans function in a weighted space for this stability analysis.
    Zeros of the Evans function indicate the intersection of the
    stable and unstable manifolds; they are used to determine the
    eigenvalues.} Within this \rbf{Evans function} framework, a
  numerical dynamical systems method for the calculation of the
  dispersion relation as an eigenvalue problem is defined and
  dispersion curves for different values of the electron diffusion
  constant and of the electric field ahead of the front are derived.
  Numerical solutions of the initial value problem confirm the
  eigenvalue calculations. The numerical work is complemented with
  \rbf{an analysis of the Evans function leading to} analytical
  expressions for the dispersion relation in the limit of small and
  large wave numbers. \rbf{The paper concludes} with a fit formula for
  intermediate wave numbers. This empirical fit supports the
  conjecture that the smallest unstable wave length of the Laplacian
  instability is proportional to the diffusion length that
  characterizes the leading edge of the pulled ionization front.


\end{abstract}

\pagestyle{myheadings} \thispagestyle{plain} \markboth{G. Derks, U.
  Ebert and B. Meulenbroek} {Laplacian instability of pulled streamer ionization fronts}

\section{Introduction}\label{sec.intro}

\subsection{The streamer phenomenon, ionization fronts and Laplacian
instability}

A streamer is the first stage of electric breakdown in large volumes, it \rbf{paves} the way of sparks and
lightning, but also occurs without successive breakdown in phenomena like sprite discharges above
thunderclouds or in corona discharges used in numerous technical applications. Recent reviews of relevant
phenomena can be found in \cite{PSST,Starikovskaia}. Considered as a nonlinear phenomenon, the streamer is a
finger shaped ionized region that propagates by self generated field enhancement at its tip into nonionized
media.  It has multiple scales as described in \cite{PSST}; as a consequence one can investigate a hierarchy
of models on different levels of refinement that are reductions of each other, starting from the reduction
from a particle to a continuum model \cite{Chao} to the reduction from a continuum model to a moving boundary
model \cite{MBA} up to the formulation of effective models for complete multiple branched streamer trees
without inner structure that are known as ``dielectric breakdown models'' \cite{DBM1,DBM2,DBM3,DBM4}. All
these reductions are \rrbf{the} subject of current research; the present paper \rbf{analyzes the stability of fronts in
the continuum model; the resulting dispersion relation provides a test case for moving boundary
approximations}.

Specifically, simulations of the simplest continuum model for negative streamers \cite{Dhali1,Dhali2,Vit}
have established the formation of a thin boundary layer around the streamer head. This layer is an ionization
front that also carries a net negative electric charge. (Positive streamers with positive net charge occur as
well, but are not \rrbf{the} subject of the present study.) The configuration of the charge in a thin layer leads to the
above mentioned field enhancement at the streamer head that creates high ionization rates and electron drift
velocities and hence lets the streamer rapidly penetrate the non-ionized region. More recent numerical
investigations show that the boundary layer or front can undergo a
Laplacian instability that \rrbf{generates} 
the
streamer branch \cite{PRL02,Andrea,Montijn1,Montijn}.  (We remark that an additional interaction mechanism in
\rbf{composite} gases like air somewhat modifies this picture \cite{Luque} while the present analysis applies
to negative streamers in simple gases like pure nitrogen or argon.)

\subsection{Moving boundary layers and the transversal instability of
pulled fronts}

The streamer can be considered as a phenomenon where an ionized phase is separated from a non-ionized phase
by a moving thin front. This concept \cite{UWC1,PRL02} implies that streamers show similar features as moving
boundary problems like viscous fingers, solidification fronts propagating into undercooled liquids, growth of
bacterial colonies or corals in a diffusive field of food etc. Quantitative predictions within such models
require a proper understanding of the front dynamics, in particular, of their \rbf{stability} against
perturbations in the transversal direction. This \rbf{stability} determines whether perturbations of the
front position will grow or shrink, and on the long term whether the streamer will branch or not. As a first
insight, one would therefore like to analyze the stability of planar fronts against transversal
perturbations, more specifically, the growth or shrinking rate $s(k)$ of a linear perturbation with
transversal wave length $2\pi/k$.

The ionization front in the model for a negative streamer in a pure
gas as treated in
\cite{Dhali1,Dhali2,Vit,UWC1,UWC2,PRL02,Andrea,Montijn1,Montijn},
including electron diffusion, creates a so-called pulled front that
has a number of peculiar mathematical properties: (\emph{i}) for each
velocity $v\ge v^*$, there is a dynamically stable front solution
where the stability is conditional on the spatial decay of the
perturbation, hence the long time dynamics is selected by the spatial
decay of the initial front for $z\to\infty$ \rrbf{(where~$z$ is the
  spatial variable along the front)}; (\emph{ii}) the convergence
towards this front is algebraically slow in time
\cite{pulled1a,pulled1b}; (\emph{iii}) this slow dynamics is
determined in the leading edge of the front that in principle extends
up to $z\to\infty$ and in the dynamically relevant space it will cause
Fredholm integrals in the linear stability analysis to diverge,
therefore curvature corrections cannot be calculated in the
established manner~\cite{pulled2}, (\emph{iv}) the unconventional
location of the dynamically relevant region ahead of the front also
requires particular care in numerical solutions with adaptive grid
refinement~\cite{Montijn}.  For the calculation of the dispersion
relation, which can be phrased as an eigenvalue problem for~$s(k)$,
these features pose two challenges: first, the condition on the
one-dimensional dynamical stability and algebraic convergence
properties, which are typical for pulled fronts, will lead to an
{apparently} degenerate eigenvalue problem. Second, in a neighborhood
of the origin, the dispersion curve~$s(k)$ is near the continuous
spectrum.  Hence numerical calculations of the eigenvalue problem with
finite difference, collocation or spectral methods often lead to
spurious eigenvalues. A dynamical systems method involving stable and
unstable manifolds avoids this problem.  \rrbf{
  The stable and unstable manifolds are {at least two-dimensional} and
  an exterior algebra approach is employed to calculate the manifolds
  accurately.}

In~\cite{Springer01,PRL02,D=0}, the treatment of pulled fronts and more-dimensional stable/unstable manifolds
\rbf{was} circumvented by neglecting the electron diffusion that acts as a singular perturbation. In this
way, the leading edge of the front together with its mathematical challenges is removed and the eigenvalue
problem can be solved using shooting on the one-dimensional stable/unstable manifolds. The resulting problem
is characterized by two length scales, namely the length scale $2\pi/k$ of the transversal perturbation, and
the longitudinal length scale of electric screening through the front that will be denoted by $\ell_\alpha$.
\rrbf{The dispersion relation in this case}
shows a quite unconventional behavior, namely a short wave length
instability whose consequences are further investigated in \cite{Bernard2,arxiv}.  In the present paper, we
analyze the dispersion relation including diffusion, mastering the above challenges and deriving quantitative
results through a combination of analytical and numerical methods.

\subsection{\rul{The Evans function and pulled fronts}}

\rbf{%
The Evans function is an analytic function whose zeros
correspond to the eigenvalues of a spectral problem, usually a linearization about a coherent structure like
a front or solitary wave. It was first introduced in~\cite{evans75} and generalized
in~\cite{agj90}. In the last decade, the Evans function has been applied in the context of many problems and
various extensions and generalizations have been found, see the review papers~\cite{kapitula04, sandstede02}
and references in there. \rrbf{One of the first uses of the Evans function
in the analysis of a planar front can be found in~\cite{terman},
which analyzes the stability of a planar wave in a reaction diffusion system
arising in a combustion model}. In
\rrbf{the current} paper we will show how pulled fronts can be analyzed with the Evans function
by using weighted spaces in its definition.}

\rbf{%
To define the Evans function, one writes the eigenvalue problem as a linear, first order, dynamical system
with respect to the spatial variable~$z$.  Along the dispersion curve $s(k)$, the dynamical system has a
solution which is bounded for all values of~$z$. This can be phrased in a more dynamical way as: the manifold
of solutions which are exponentially decaying for $z\to+\infty$ (stable manifold) and the manifold of
solutions which are exponentially decaying for $z\to-\infty$ (unstable manifold) have a non-trivial
intersection along the dispersion curve. The Evans function is a function of the spectral parameters $s$ and
$k$, which vanishes if the stable and unstable manifolds have a non-trivial intersection.  Hence the Evans
function can be viewed as a Melnikov function or a Wronskian determinant, see also~\cite{kapitula99} or
references in there.  }

\rbf{%
In case of a pulled front, the definition of the stable manifold, and hence the Evans function, is not
straightforward. The temporal stability of the asymptotic state of the pulled front at $+\infty$ is
conditional on the spatial decay of the perturbation. So this decay condition should be included in the
definition of the stable manifold, otherwise the dimension of this manifold might be too large.
We will show that this condition can be built in the definition of the stable manifold by considering the
stable manifold in a weighted space. The Evans function is defined by using the weighted space for the stable
manifold. Hence the dispersion curve~$s(k)$ can be found as a curve of zeros of this Evans function. }

\subsection{\rul{Organization of the paper}}

In section~\ref{sec.2}, we recall the model equations and the construction and properties of planar fronts.
In particular, we summarize the multiplicity, stability, dynamical selection and convergence rate of these
pulled fronts.
In section~\ref{sec.3}, the 
\rbf{stability} of these fronts is investigated as an eigenvalue problem for the dispersion relation~$s(k)$
of a linear perturbation with wave number~$k$.  The dispersion relation depends on the far electric
field~$E_\infty$ and the electron diffusion~$D$ as external parameters.  In the stability analysis of the
pulled ionization fronts, a constraint is imposed on the asymptotic spatial decay rate of the perturbations.
This constraint corresponds to the decay condition for the one-dimensional stability, but has to be chosen
quite subtly to avoid problems with the algebraic decay of the front solution. A consequence of the decay
condition is that the eigenvalue problem (dispersion relation) is solved in a weighted space. In this
weighted space, the {apparent} degeneracies have disappeared, the stable and unstable manifolds of the ODE
related to the eigenvalue problem are well-defined and intersections of those manifolds are determined by
using the Evans function.
In section~\ref{sec.results}, dispersion relations for positive~$s$ are derived numerically for a number of
pairs of external parameters $(E_\infty,D)$. The numerical implementation of the Evans function uses exterior
algebra to reliably solve for the \rbf{higher dimensional} stable and unstable manifolds.
In section~\ref{sec.simulation}, the numerical dispersion relation is tested \rbf{thoroughly} and confirmed
with numerical simulations of the initial value problem for the complete PDE model \rbf{for the particular
values $(E_\infty,D)=(-1,0.1)$} \rbf{ where $D=0.1$ is typically used for nitrogen
\cite{Dhali1,Dhali2,Vit,UWC1,UWC2,PRL02,Andrea,Montijn1,Montijn} and $E_\infty=-1$ is a representative value
for the electric field. The later sections treat either general $(E_\infty,D)$ analytically or a larger range
of $(E_\infty,D)$ numerically.}

In section~\ref{asymptotics}, explicit analytical 
\rbf{asymptotic relations} for the dispersion relation~$s(k)$ are
derived for the limits of small and large wave numbers~$k$. 
\rbf{For $k=0$, two explicit eigenfunctions are known (which are related to the
translation and gauge symmetry in the problem). These explicit solutions
lead to expressions for the solutions on the stable manifold for
small wave numbers.}
The interaction between the slow and fast
behavior on this manifold leads to an asymptotic dispersion relation
for small~$k$.  For large wave numbers, the eigenvalue problem for the
dispersion relation is dominated by a constant coefficient eigenvalue
problem.  \rrbf{An eigenvalue exists only if this constant coefficient
  system has no spectral gap}. Using exponential dichotomies and the
roughness theorem, the asymptotics of the dispersion relation is
derived by a contradiction argument.
In section~\ref{sec.6}, these \rbf{asymptotic limits} are tested on the numerical data derived in
section~\ref{sec.3}.  It is found that the 
\rbf{asymptotic limit} for small $k$
fits the data very well, while the 
\rbf{asymptotic limit} for large $k$ is not yet applicable in the range where $s(k)$ is positive. After a
discussion of relevant physical scales, we suggest a fit formula joining the
analytical small $k$ 
\rbf{asymptotic limit} with our physically motivated guess. This formula fits the numerical data well for
practical purposes and strongly supports the conjecture that the smallest unstable wave length is
proportional to the diffusion length that determines the leading edge of the pulled front.

\section{The streamer model and its ionization fronts}\label{sec.2}

In this section we describe the streamer model and summarize the features of planar ionization fronts as
solutions of the purely one-dimensional model as a preparation for the stability analysis in the {dimensions
transversal to the front}. In particular, we recall the multiplicity of the front solutions that penetrate a
linearly dynamically unstable state, and the dynamical selection of the pulled front.

\subsection{Model equations}

We investigate negative fronts within the minimal strea\-mer model,
i.e., within a ``fluid approximation'' with local field-dependent
impact ionization reaction in a non-attaching gas like argon or
nitrogen \cite{UWC1,UWC2,Springer01,PRL02,Andrea}. The equations for
this model in dimensionless quantities are
\begin{eqnarray}
\label{107}
\partial_t\;\sigma \;-\; D\nabla^2\sigma
\;-\;\nabla\cdot\left(\sigma \;{\bf E}\right) &=& \sigma \; f(|{\bf E}|)~,
\\
\label{108}
\partial_t\;\rho
&=& \sigma \; f(|{\bf E}|)~,
\\
\label{109}
\nabla\cdot{\bf E} &=& \rho-\sigma~,~~{\bf E}=-\nabla \phi~,
\end{eqnarray}
where $\sigma$ is the electron and $\rho$ the ion density, {\bf E} is
the electric field and $\phi$ is the electrostatic potential.  For
physical parameters and dimensional analysis, we refer to discussions
in \cite{UWC1,UWC2,Springer01,PRL02,Andrea}.  The electron current is
approximated by diffusion and advection $-D\nabla\sigma - \sigma {\bf
  E}$. The ion current is neglected, because the front dynamics takes
place on the fast time scale of the electrons and the ion mobility is
much smaller.  Electron--ion pairs are assumed to be generated with
rate $\sigma f(|{\bf E}|) = \sigma|{\bf E}|\alpha(|{\bf E}|)$ where
$\sigma|{\bf E}|$ is the absolute value of electron drift current and
$\alpha(|{\bf E}|)$ the effective impact ionization cross section within
a field ${\bf E}$. Hence $f(|{\bf E}|)$ is
\be \label{f} f(|{\bf E}|)=|{\bf E}|\;\alpha(|{\bf E}|)~. \ee
For numerical calculations, we use the Townsend approximation $\alpha(|{\bf E}|)=e^{-1/|{\bf
E}|}$~\rbf{\cite{UWC1,UWC2,Springer01,PRL02,Andrea}}.  For analytical calculations, an arbitrary function
$\alpha({|\bf E}|)\ge 0$ can be chosen where we assume that $\alpha(0)=0$ and therefore $f(0)=0=f'(0)$. We
will furthermore assume that $\alpha(|{\bf E}|)$ is monotonically increasing in $|{\bf E}|$, this is a
sufficient criterion for the front to be a pulled one~\cite{pulled1b}. The electric field can be calculated
in electrostatic approximation ${\bf E}=-\nabla\phi$.

Mathematically, the model~\eqref{107}-\eqref{109} describes the
dynamics of the three scalar fields $\sigma$, $\rho$ and $\phi$. It is
a set of reaction-advection-diffusion equations for the charged
species $\sigma$ and $\rho$ coupled nonlinearly to the Poisson
equation of electrostatics.

\subsection{Two types of stationary states}

It follows immediately from~\eqref{107}-\eqref{109} that there can be two types of stationary states of the
system, one characterized by $\sigma\equiv0$ and the other by ${\bf
  E}\equiv0$ 
\rbf{(as $f(|{\bf E}|)=0$ implies $|{\bf E}|=0$.)}.

The stationary state with $\sigma\equiv0$ is the non-ionized state. As the dynamics is only carried by the
electrons $\sigma$, there is no temporal evolution for $\sigma\equiv0$ even if the ion density $\rho$ has
\rbf{an arbitrary spatial distribution}. The electric field ${\bf E}=-\nabla\phi$ then is determined by the
solution of the Poisson equation $-\nabla^2\phi=\rho$ and by the boundary conditions on $\phi$. In certain
ionization fronts in semiconductor devices \cite{Pavel}, it is essential that the equivalent of $\rho$ does
not vanish in the non-ionized region. In the gas discharges considered here, on the other hand, it is
reasonable to assume that the non-ionized initial state with $\sigma\equiv0$ also has a vanishing ion density
$\rho\equiv0$, and therefore no space charges.

The stationary state with vanishing electric field ${\bf E}\equiv0$ describes the ionized, electrically
screened charge neutral plasma region behind an ionization front, the interior of the streamer. From ${\bf
E}\equiv0$ the identity $\nabla\cdot{\bf E}=0$ follows immediately, and therefore electron and ion densities
have to be equal $\sigma=\rho$. In the absence of a field, the electrons diffuse
$\partial_t\sigma=D\nabla^2\sigma$ while the ions stay put $\partial_t\rho=0$. \rbf{Therefore, these
densities only can stay equal if $\nabla^2\rho=0$.
Simulations~\cite{Dhali1,Dhali2,Vit,UWC1,UWC2,PRL02,Andrea,Montijn1,Montijn} show that this occurs typically
only if $\rho$ is homogeneous (though counter examples can be constructed).}

\subsection{Planar ionization front solutions}

An ionization front separates such different outer regions: an electron-free and non-conducting state with an
arbitrary electric field $E_\infty$ ahead of the front from an ionized and electrically screened state with
arbitrary, but equal density $\sigma^-=\rho^-$ of electrons and ions.  In particular, we are interested in
almost planar fronts propagating into a particle-free region $\rho=\sigma=0$ (where therefore
$\nabla^2\phi=0$), and we study negative fronts, i.e., fronts with an electron surplus that propagate into
the electron drift direction towards an asymptotic electric field~$E_\infty<0$. For a planar front, it
follows from $\nabla^2\phi=-\nabla\cdot{\bf E}=0$ that the electric field ahead of the front is homogeneous.

\rbf{We assume that the front propagates into the positive $z$ direction; the electric field ahead of a
negative front then is ${\bf E}\to E_\infty \hat{\bf z}$, $E_\infty<0$, for $z\to\infty$}. (Here $\hat{\bf
z}$ is the unit vector in the $z$-direction.)  It is convenient to introduce the coordinate system
$(x,y,\xi=z-vt)$ moving with the front velocity ${\bf v}=v\hat{\bf z}$. A planar, uniformly translating front
is a stationary solution in this co-moving frame, hence \rbf{it} depends only on the co-moving coordinate
$\xi$, and will be denoted by a lower index 0. A front satisfies
\begin{equation}\label{UTF}
\begin{arrayl}
D\partial_\xi^2\sigma_0 &
{} + (v-\partial_\xi\phi_0)\;\partial_\xi\sigma_0 &
{} + \sigma_0(\rho_0-\sigma_0) + \sigma_0 f_0 =0,
\\
&v\partial_\xi\rho_0 &{} + \sigma_0 f_0=0,
\\
\partial_\xi^2\phi_0&&{} + \rho_0-\sigma_0=0,
\end{arrayl}
\end{equation}
where $f_0=f(|E_0|)$.  This system can be reduced to 3 first order ordinary differential equations. First,
due to electric gauge invariance, the system does not depend on $\phi_0$ explicitly, but only on
$E_0=-\partial_\xi\phi_0$. Using the variable $E_0$ instead of $\phi_0$ reduces the number of derivatives by
one. Second, electric charge conservation $\partial_t q+\nabla\cdot {\bf j}=0$ \rbf{can be rewritten in
co-moving coordinates for a uniformly translating front as} $-v\partial_\xi q_0+\partial_\xi j_0=0$.
Therefore it can be integrated once $-vq_0+j_0=c$, $\partial_\xi c=0$.  In the present problem, the space
charge is $q_0=\rho_0-\sigma_0$ and the electric current is $j_0=-D\partial_\xi\sigma_0-\sigma_0E_0$.
Furthermore, as there is a region with vanishing densities $\sigma_0=0=\rho_0$ ahead of the front, the
integration constant $c$ vanishes in this region, and therefore everywhere. Thus the planar front
equations~(\ref{UTF}) can be written as
\be D\, \partial_\xi\sigma_0 &=& v \,
(\rho_0-\sigma_0) - E_0\,\sigma_0,
\nonumber\\
v\, \partial_\xi\rho_0 &=& - \sigma_0\, f(|E_0|),\label{eq.front}
\\
\partial_\xi E_0 &=& \rho_0-\sigma_0,
\nonumber
\ee
where $\partial_\xi\phi_0 = - E_0$ decouples from the other
equations. The planar front equations imply that $E_0(\xi)<0$ for all
$\xi$ when $E_\infty<0$~\cite{UWC2}.

The fronts connect the states
\be \label{bounds}
\left(\begin{array}{c}\sigma_0\\
    \rho_0\\ E_0\end{array}\right)
\stackrel{\xi\to+\infty}{\rightarrow}
\left(\begin{array}{c} 0\\ 0\\ E_\infty\end{array}\right)
~~~ \mbox{and}~~~
\left(\begin{array}{c} \sigma_0\\
    \rho_0\\ E_0\end{array}\right)
\stackrel{\xi\to-\infty}{\rightarrow}
\left(\begin{array}{c} \sigma^-\\
    \sigma^-\\0\end{array}\right),
\ee
and the electrostatic potential~$\phi_0$ connects $\phi^-$ (for $\xi\to-\infty$) with $-E_\infty\xi+\phi^+$
(for $\xi\to+\infty$). The ionization density $\sigma^-$ behind the front and the electrostatic potential
difference $\phi^+-\phi^-$ have to be determined for arbitrarily chosen electric field $E_\infty$ ahead of
the front and for arbitrary, but sufficiently large, front velocity $v$.  (We remark that only the potential
difference $\phi^+-\phi^-$ matters due to the gauge invariance of the electrostatic potential as one easily
verifies on the equations.)  The fronts can be constructed as heteroclinic orbits in a three-dimensional
space as demonstrated in \cite{UWC2}.

The diffusion constant $D$ is obviously a singular perturbation.  For
$D=0$, the front equations can be solved analytically \cite{UWC2,D=0},
i.e., one can find explicit expressions for the particle densities
$\sigma_0[E_0]$, $\rho_0[E_0]$ and for the front coordinate $\xi[E_0]$
as a function of the electric field~$E_0$.  For the negative fronts
treated here, 
\rbf{the front is continuous as function of $D$ and
the limit $D\to0$ exists and equals the value of the front at $D=0$}, while for
positive fronts ($E_\infty>0$), it is singular \cite{UWC2}.

\subsection{Multiplicity of front solutions, pulled fronts and dynamical selection}\label{sec.mult}

The non-ionized state $(\sigma,\rho,E)=(0,0,E_\infty)$ with a nonvanishing electric field $E_\infty$ is
linearly unstable under the temporal dynamics of the PDE~\eqref{107}-\eqref{109}. In fact, this spatial
region ahead of the front dominates the dynamics, cf.\ the discussion in \cite{UWC2,pulled1b}. Therefore, for
fixed~$E_\infty$, there is a continuous family of
uniformly translating solutions, parametrized by the velocity $v\ge v^*$~\cite{UWC1,UWC2,pulled1a,pulled1b},
where
\be \label{v*} v^*(E_\infty)=|E_\infty|+2\sqrt{D\;f(|E_\infty|)}. \ee
The dynamics of uniformly translating fronts with velocity $v>v^*$ are dominated by a flat spatial profile in
the leading edge of the \rbf{front
\be \label{flat} \sigma_v(\xi)\stackrel{\xi\to\infty}\sim e^{-\lambda  \xi} ~~~\mbox{with }
\lambda<\Lambda^*=\sqrt{\frac{f(|E_\infty|)}{D}}, \ee
where velocity $v$ and decay rate $\lambda$ are related through
\be v(E_\infty,\lambda)= |E_\infty|+D\lambda+\frac{f(E_\infty)}\lambda, \ee
and therefore $v(E_\infty,\lambda)>v^*(E_\infty)\equiv v(E_\infty,\Lambda^*)$ for $\lambda\ne\Lambda^*$. The
spatial profile (\ref{flat}) with $\lambda<\Lambda^*$ cannot build up dynamically from some initial condition
with larger $\lambda$; and it will destabilize if perturbed with an initial condition with smaller $\lambda$,
therefore such flat and fast fronts can be approached dynamically only by initial conditions with exactly the
same profile (\ref{flat}) in the leading edge.
For a thorough discussion of this dynamics, we refer
to~\cite{pulled1b}}.

In practice, the continuum approximation for the electron density
breaks down for very small densities in the leading edge and
\rbf{the initial electron distribution satisfies} a decay
condition of the form
\be \label{InitDecay}
\lim_{\xi\to\infty} \sigma(x,y,\xi,t=0)\;e^{\lambda \xi}=0
\qmbox{for all} \lambda < \Lambda^* , 
\ee
if the penetrated state is really non-ionized.  Therefore the velocity $v^*$ is called the
``selected'' one, because it is the generic attractor for most physical initial conditions. Mathematically
speaking, the profile with velocity~$v^*$ (the selected front) is the only profile that can build up
dynamically from steeper initial conditions.

Therefore the condition~\eqref{InitDecay} on the spatial decay of the initial electron distribution excludes
all front solutions with velocity $v>v^*$ as long time attractors of the dynamics.  If the
criterion~\eqref{InitDecay} is satisfied, then the selected front with speed~$v^*$ is dynamically stable and
is approached with the universal algebraic convergence rate in time \cite{pulled1a,pulled1b}
\be \label{relax}
v(t)=v^*-\frac3{2 \Lambda^* t} +{\cal O}\left(\frac 1 {t^{3/2}}\right). \ee
However, without the spatial decay condition on the initial condition, the selected front is formally not
stable (although this is physically irrelevant). This will lead to specific problems and solutions in
the transverse stability analysis presented in the next section.

The spatial profile of the electron distribution in the selected front is
\be \label{Profile}
\sigma_{v^*}(\xi)\stackrel{\xi\to\infty}\sim
(a\xi+b)~e^{-\Lambda^*\xi},~~~a>0.
\ee
To summarize, if the analysis is restricted to initial conditions with a sufficiently rapid spatial decay in
the electron densities (\ref{InitDecay}), then the fronts have only one free external parameter, namely the
field $E_\infty$; it determines the asymptotic front velocity~\eqref{v*} and profile~(\ref{Profile}) after
sufficiently long times.  Furthermore, the equivariance in the system gives that the position of the front
and its electrostatic potential are free internal parameters.

\subsection{Full spatial profiles of the selected pulled planar front}

The spatial decay behind the front will be important in the analysis as well, therefore we recall the basic
behavior. For $\xi\to-\infty$, the electron density approaches
\be \label{Profile-} \sigma_{v^*}(\xi)\stackrel{\xi\to-\infty}=\sigma^- + c~e^{\lambda^-\xi},~~~c>0, \ee
and the electric field decays with the same exponent $E(\xi)=-(c/\lambda^-)~e^{\lambda^-\xi}$.
For $D=0$,
\be \label{sigma-0}\sigma^-(E_\infty,D=0)=\int_0^{|E_\infty|}\alpha(x)\;dx \ee
was derived in \cite{UWC2}. For $D>0$, $\sigma^-$ \rbf{decreases by a correction of order of $D$, more
precisely,}
\be \label{sigma-D} \sigma^-(E_\infty,D)=\sigma^-(E_\infty,0)+{\cal O}(D), \quad
\sigma^-(E_\infty,D>0)<\sigma^-(E_\infty,0) \ee
was proved in the appendix of \cite{Chao}. The eigenvalue $\lambda^-$ is given by
\be \label{lambda-}\lambda^-=\frac{\sqrt{v^{*2}+4D\sigma^-}-v^*}{2D}, \ee
where both $v^*$ and $\sigma^-$ depend on $E_\infty$ and $D$. For small $D$, $\lambda^-$ can be approximated
as
\be \label{lambda-0} \lambda^- = \frac{\sigma^-}{v^*}+{\cal O}(D) = \int_0^{|E_\infty|}
\frac{\alpha(x)\;dx}{|E_\infty|}+{\cal O}(\sqrt{D}) . \ee

\begin{figure}[htb]
  \centering
  \includegraphics[width=0.483\textwidth]{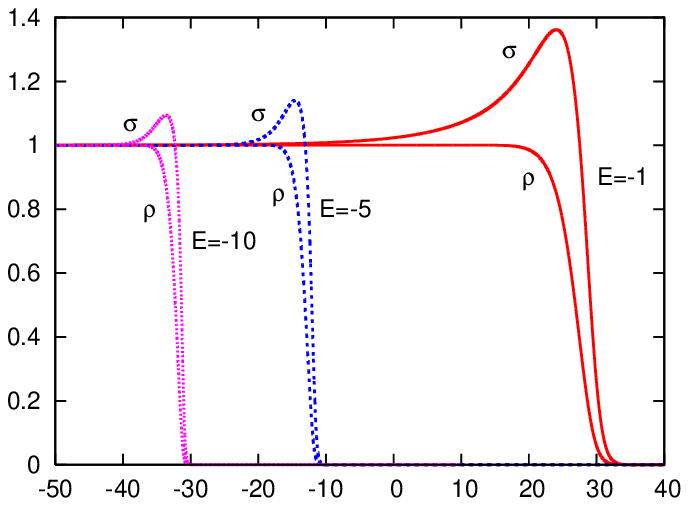}\quad
  \includegraphics[width=0.483\textwidth]{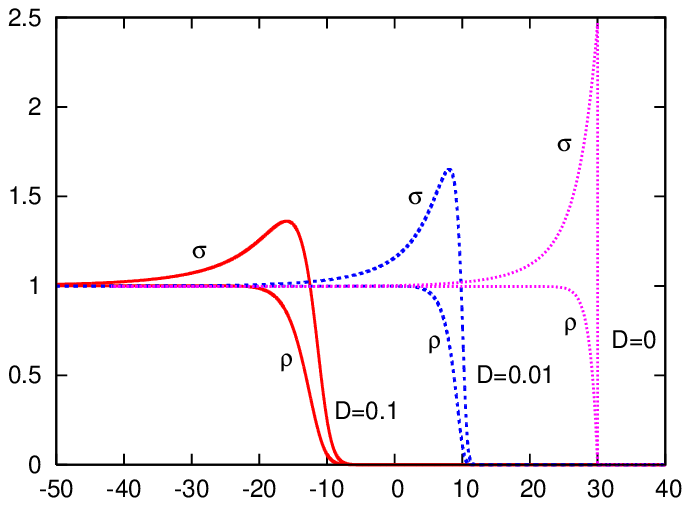} \\
  \centering
  \includegraphics[width=0.483\textwidth]{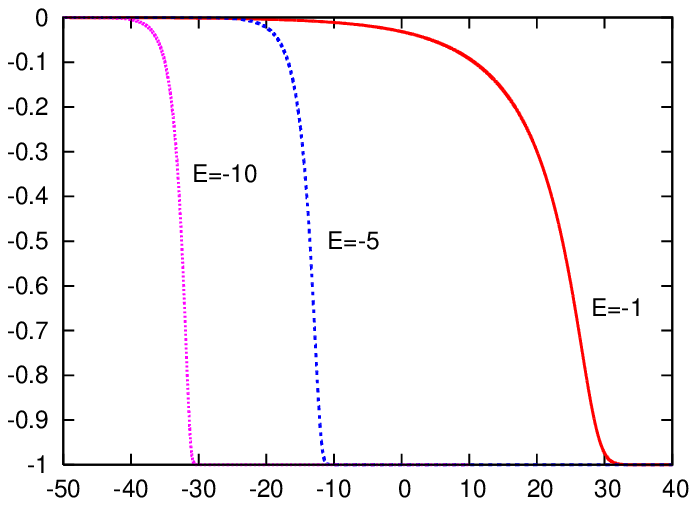}\quad
  \includegraphics[width=0.483\textwidth]{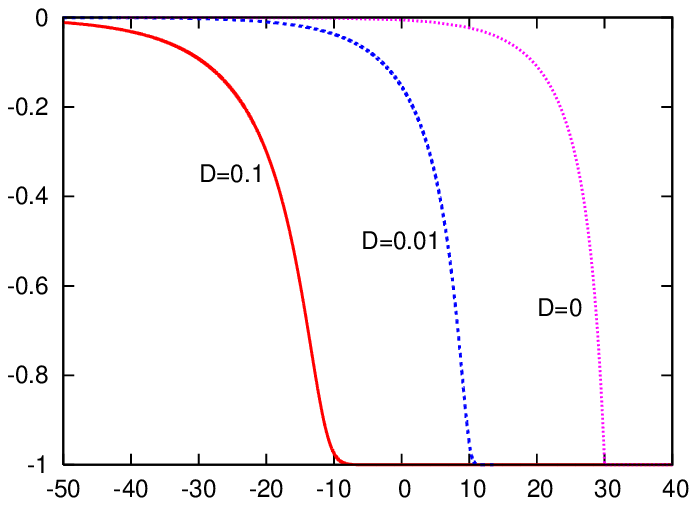}
\caption{The pulled planar front solutions on the left for varying $E_\infty = -1$,
  $-5$ and $-10$ and fixed $D=0.1$, and on the right for fixed $E_\infty=-1$
  and varying $D=0.1$, $0.01$ and $0$. The upper panels show scaled electron
  and ion densities $\sigma_0(\xi)/\sigma^-(E_\infty,D)$ and
  $\rho_0(\xi)/\sigma^-(E_\infty,D)$ and the lower panels the corresponding
  scaled electric fields
  $E_0(\xi)/|E_\infty|$ as a function of the spatial coordinate $\xi$.
  The fronts are displayed in a staggered way. The normalization factors
  $\sigma^-(E_\infty,D)$ in the upper panels are $\sigma^-(-1,\;0)=0.149$,
  $\sigma^-(-1,\;0.01)=0.148$, $\sigma^-(-1,\;0.1)=0.144$,
  $\sigma^-(-5,\;0.1)=2.832$, $\sigma^-(-10,\;0.1)=7.169$.}
  \label{fig.front}
\end{figure}

As an illustration, the spatial profiles of electron and ion density and the electric field of the selected
front solution for a range of fields $E_\infty$ and diffusion constants $D$ are plotted in
Figure~\ref{fig.front}.

\section{Numerical calculation of the dispersion relation}\label{sec.3}

First we will introduce the transversal perturbation setting and
discuss an apparent degeneracy of the dispersion relation. However, it
turns out that the constraint on the spatial decay of the electron
density ``selects'' a single dispersion relation for every far field
$E_\infty$.  This relation then is calculated numerically based on
dynamical systems techniques involving intersections of stable and
unstable manifolds.  Results for different fields $E_\infty$ and
diffusion constants $D$ are presented.

\subsection{Linear transversal perturbations of planar fronts}

Suppose that there is a linear transversal perturbation of the
uniformly translating front
\be \sigma(x,y,\xi,t)= \sigma_0(\xi)
+\delta\;\overline{\sigma}_1(x,y,\xi,t)+O(\delta^2),~~~ \xi=z-v^*t, \ee
and similarly for $\rho$ and $\phi$.  The linearized equation for
$\overline{\sigma}_1$, $\overline{\rho}_1$ and $\overline{\phi}_1$
follows from Eqs.~\eqref{107}-\eqref{109}.  By decomposing the
perturbations into Fourier modes in the transversal directions $x$ and
$y$, by using isotropy in the transversal $(x,y)$ plane and by using a
Laplace transformation in $t$, the ansatz
\be
(\overline{\sigma}_1,\overline{\rho}_1,\overline{\phi}_1)=e^{ikx+st}
\,(\sigma_k,\rho_k,\phi_k)(\xi) \ee
can be used for each Fourier component.
\rbf{The resulting equation can be written as a linear first order system
  of ODEs, using the extra variables $\tau_k=\partial_\xi \sigma_k$
  and $E_k=-\partial_\xi \phi_k$.  Introduce ${\bf w} =
  (\tau_k,\sigma_k,\rho_k,E_k,\phi_k)$ and the linear system is given by}
\be \label{Pert5E} \label{eq.linsys}
&&\partial_\xi {\bf w} = {\bf M}(\xi;E_\infty,k,s)\, {\bf w},
\\[2mm]&&
\mbox{ with }
{\bf M} =
\begin{pmatrix}
  -\frac{E_0+v^*}{D} & \frac{2\sigma_0-\rho_0-f_0+s+Dk^2}D &
   -\frac{\sigma_0}D & -\frac{\partial_\xi\sigma_0-\sigma_0 f'_0}D & 0\\[1mm]
1 & 0&0&0&0\\[1mm]
0&-\frac{f_0}{v^*} &\frac s {v^*} &  \frac{\sigma_0 f'_0}{v^*} &
  0\\[1mm]
0&-1&1&0&-k^2\\[1mm]
0&0&0&-1&0
\end{pmatrix}.
\nonumber
\ee
\rbf{In the matrix~$M$}, the abbreviated notations $f_0=f(|E_0|)$ and
$f'_0=\partial_\eta f(\eta)\big|_{\eta=|E_0|}$ are used. \rbf{For} the terms
with $f_0'$, we have used that $E_0<0$, hence $\frac{E_0}{|E_0|} =-1$.

As the matrix $\bM$ depends on $k^2$, but not on $k$ itself, the
matrix is invariant under the transformation $k \to -k$. Thus if
$s(k)=s^*$, then also $s(-k)=s^*$ and vice versa. Therefore it is
sufficient to determine the dispersion relation for $k>0$ and this
will imply the relation for $k<0$ and from now on, we will use the
convention that $k>0$. \rrbf{Note that the invariance implies only
  that the dispersion relation will be a function of $|k|$. As will be
  shown later, the dispersion relation is not an analytic function
  of~$k$ near $k=0$ and its expansion near $k=0$ is linear in $|k|$.}

For future use, we remark that the linearization matrix~$\bM$ does not
involve any $\xi$-dependent terms in the fourth and fifth row and
implies that $E_k$ and $\phi_k$ are related by $E_k'=-\phi_k$. Thus
the $E_k$-component of any solution of the linearized
system~\eqref{eq.linsys} can be expressed as an integral
\begin{equation}\label{eq.E_implicit}
E_k(\xi) = c_1 e^{k\xi} + c_2 e^{-k\xi} + \frac 12 \int_{\xi_0}^\xi
\left[e^{k(\xi-\eta)}+e^{-k(\xi-\eta)}\right]
\left[\rho_k(\eta)-\sigma_k(\eta)\right]\, d\eta,
\end{equation}
where the constants $c_1$ and $c_2$ are determined by the value of
$E_k$ and $\phi_k$ at $\xi=\xi_0$.

\subsection{Stable and unstable manifolds and degeneracy of the
dispersion relation}
\label{sec.dyn_sys}

The linearized problem~\eqref{Pert5E} is a spectral problem with the
spectral parameters~$s$ and~$k$.  If the asymptotic matrices
$\bM^\pm(E_\infty,k,s) = \lim_{\xi\to\pm\infty} \bM(\xi;E_\infty,k,s)$
exist and are hyperbolic (i.e., no eigenvalues on the imaginary axis),
then the system~\eqref{Pert5E} has a bounded solution if and only if
the unstable manifold from $\xi=-\infty$ and the stable manifold from
$\xi=\infty$ have a non-trivial intersection.  So we will focus in
this section on determining the stable and unstable manifolds.

The behavior of the unstable manifold at the back of the front is
given by the asymptotic matrix
\[
{\bf M}_-(E_\infty,k,s) = \mijnlim{\xi\to -\infty} \bM(\xi;E_\infty,k,s) =
\begin{pmatrix}
  -\frac{v^*}{D} & \frac{\sigma^- + s + Dk^2}D &
   -\frac{\sigma^-}D & 0 & 0\\[1mm]
1 & 0&0&0&0\\[1mm]
0& 0 &\frac s {v^*} & 0 & 0\\[1mm]
0&-1&1&0&-k^2\\[1mm]
0&0&0&-1&0
\end{pmatrix}.
\]
For $s>0$ and $k\neq0$, this matrix has two negative and three
positive eigenvalues:
\begin{equation}\label{eq.eigenval-}
\pm k , \quad \frac s {v^*} , \quad \mu^-_\pm=-\frac {v^*}{2D} \pm
 \frac{\sqrt{v^{*2}+4D(\sigma^-+s+Dk^2)}}{2D}.
\end{equation}
Thus the unstable manifold is three dimensional. {We remark that $\mu_+^-(s=0=k)$ is identical to the
spatial decay rate $\lambda^-$ (\ref{lambda-}) behind the unperturbed front.}

Finding the stable manifold ahead of the front is less straightforward. Normally the stable manifold ahead of
the front would be characterized by the matrix $\lim_{\xi\to +\infty} \bM(\xi;E_\infty,k,s)$. For $s>0$ and
$s+Dk^2<f(E_\infty)$ this matrix exists and has two positive and three negative eigenvalues:
\be \label{EW} \pm k,\quad \frac s {v^*}, \quad
-\Lambda^*\pm\sqrt{\frac{s+Dk^2}D} =
\frac{-\sqrt{f(E_\infty)}\pm\sqrt{s+Dk^2}}{\sqrt{D}},
\ee
\rbf{Thus the stable manifold is three dimensional and a dimension count gives that the intersection of stable
and unstable manifold is generically one dimensional. So for small values of $s$ and $k$, a continuous family
of eigenvalues seems to exist. This feature is related to the instability of the asymptotic state at
$+\infty$, to the continuous family of uniformly translating solutions for all $v\ge v^*(E_\infty)$, and to
the instability of fronts against perturbations with smaller spatial decay rate $\lambda$, as discussed in
the previous section. The continuous family of eigenvalues $s$ for fixed wave number $k$ is eliminated by
applying the analysis only to fronts with a sufficiently rapid spatial decay~\eqref{InitDecay}. This
condition will be imposed in the definition of the stable manifold; it will make the spectrum discrete.}

Define the scaled vector
\begin{equation}\label{eq.scaling}
\widetilde{\bf w} = {\bf D}\, {\bf w},~~~
{\bf D} = \diag (e^{(\Lambda^*-\beta)\xi}, e^{(\Lambda^*-\beta) \xi},
1,1,1)
\end{equation}
where $\beta\in(0,\Lambda^*)$ will be fixed later and depend on $k$ and
$\Lambda^*$. The freedom in the choice of $\beta$ is reminiscent of the
fact that the decay condition holds for any $\lambda<\Lambda^*$, but
not for $\lambda=\Lambda^*$.  The system for $\widetilde{\bf w}$ is
\begin{equation}\label{eq.linsys_scaled}
\widetilde{\bf w}_\xi = \widetilde \bM (\xi;E_\infty,k,s)\, \widetilde{\bf w},
\end{equation}
with
\begin{eqnarray}
&&\widetilde \bM = {\bf D}\cdot\bM\cdot{\bf D}^{-1} +
(\partial_\xi{\bf D})\cdot{\bf D}^{-1}
\nonumber\\
&& =\begin{pmatrix}
  -\frac{E_0+v^*}{D} +\Lambda^*-\beta& \frac{2\sigma_0-\rho_0-f_0+s+Dk^2}D &
  -\frac{\sigma_0}D \, e^{(\Lambda^*-\beta)\xi}&
  -\frac{\partial_\xi\sigma_0-\sigma_0f'_0}D \,
  e^{(\Lambda^*-\beta)\xi}& 0\\[1mm]
  1 & \Lambda^*-\beta&0&0&0\\[1mm]
  0&-\frac{f_0}{v^*}\,e^{-(\Lambda^*-\beta)\xi} &\frac s {v^*}&
   - \frac{-\sigma_0f'_0}{v^*}& 0\\[1mm]
  0&-e^{-(\Lambda^*-\beta)\xi}& 1 &
  0&-k^2\\[1mm]
  0&0&0&-1&0
\end{pmatrix}
\nonumber
\end{eqnarray}

Note that if $\beta=0$, then the asymptotic matrix ahead of the front (at $\xi=+\infty$) does not exist
because $e^{\Lambda^* \xi}\sigma_0(\xi)$ grows linearly in $\xi$ according to~\eqref{Profile}. To get an
asymptotic matrix ahead of the front, it is necessary that $0<\beta<\Lambda^*$. In this case, the asymptotic
matrix is
\[
\widetilde{\bf M}_+(E_\infty, k,s) =
\mijnlim{\xi\to\infty} \widetilde\bM(\xi;E_\infty,k,s) =
\begin{pmatrix}
  -\Lambda^*-\beta& \frac{-f_\infty+s+Dk^2}D &
  0 & 0 & 0\\[1mm]
  1 & \Lambda^*-\beta&0&0&0\\[1mm]
  0&-\frac{f_\infty}{v^*} &\frac s {v^*} &
   0 & 0\\[1mm]
  0&0&0&0&-k^2\\[1mm]
  0&0&0&-1&0
\end{pmatrix}
\]
where $f_\infty=f(|E_\infty|)$.
The matrix~$\widetilde\bM_+$ has the eigenvalues
\begin{equation}\label{eq.eigenval+}
\pm k , \quad \frac s {v^*} , \qmbox{and} \mu^+_\pm=
-\beta \pm\sqrt{\frac{s + Dk^2}{D}}.
\end{equation}
Hence for $s>0$ and $0<\beta<\min(\Lambda^*,k\sqrt{1+s/(Dk^2)})$,
there are two negative and three positive eigenvalues. Thus the stable
manifold of~\eqref{eq.linsys_scaled} is two dimensional.

For the original unscaled system~\eqref{eq.linsys} this means that
only the two-dimensional submanifold given by ${\bf D}^{-1}$ acting on
the stable manifold of~\eqref{eq.linsys_scaled} is relevant for the
transverse instability.  This submanifold will be called the stable
manifold of~\eqref{eq.linsys} from now on.  With this convention, the
dispersion relation is a well-defined curve~$s(k)$ and the curve is
such that at $s=s(k)$, the linearized system~\eqref{eq.linsys} has a
bounded solution which satisfies the spatial decay
condition~\eqref{InitDecay}. 
\rrbf{Note that for both asymptotic matrices~$\widetilde\bM_+$ and~${\bf
    M}_-$, the eigenvalues $\pm k$ become a degenerate eigenvalue~0 at
  $k=0$. This leads to square root singularities and it can be
  expected that the dispersion relation $s(k)$ will be a function of
  $\sqrt{k^2}=|k|$ for $k$ is small. This will be confirmed in
  section~\ref{asymptotics}. }

\subsection{The Evans function for the transverse stability problem}

The occurrence of an intersection of the stable and unstable manifolds
\rbf{will} be measured \rbf{with} the Evans function.  Our numerical
method to determine the dispersion curve as an eigenvalue problem is
based on \rbf{a definition of} the Evans function in an exterior
algebra framework and uses similar ideas as
in~\cite{ab02,brin,bz02,blyuss03,derks02,derks05}.  The approach of
following the stable/unstable manifolds at $\xi=\pm \infty$ with a
standard shooting method and checking their intersection using the
Evans function, works only if these manifolds are one-dimensional or
have co-dimension one; in the present model, this is the case in the
singular limit $D=0$ and a shooting method was used in~\cite{D=0}.
Otherwise, any integration scheme will inevitably just be attracted by
the eigendirection corresponding to the most unstable (stable)
eigenvalue.  Exterior algebra can be used to avoid this problem for
higher dimensional manifolds and to preserve the analytic properties
of the Evans function. \rrbf{Recently, a different method to calculate
  the Evans function for higher dimensional manifolds has been
  proposed in~\cite{humpherys06a}.  This method uses a polar
  coordinate approach and looks like a more suitable method for very
  high dimensional problems.}

To calculate the evolution of the two dimensional stable and three dimensional unstable manifold in a
reliable numerical way, we will use the exterior algebra spaces $\bigwedge^2(\mathbb{C}^5)$ \textbf{and}
$\bigwedge^3(\mathbb{C}^5)$, respectively. The advantage of these spaces is that in
$\bigwedge^l(\mathbb{C}^n)$, an $l$-dimensional linear subspace of $\mathbb{C}^n$ can be described as a
one-dimensional object, being the $l$-wedge product of a basis of this space.  Also, the differential
equation on $\mathbb{R}^5$ (or $\mathbb{C}^5$) induces a differential equation on the spaces
$\bigwedge^l(\mathbb{C}^5)$:
\begin{equation}\label{Unk}
{\bf W}_\xi = {\bf M}^{(l)} (\xi;E_\infty,k,s){\bf W} , \quad {\bf W}\in
\textstyle\bigwedge^l(\mathbb{C}^5).
\end{equation}
Here the linear operator (matrix) ${\bf M}^{(l)}$ is defined on a
decomposable $l$-form~${\bf w}_1\wedge\ldots\wedge{\bf w}_l$, $ {\bf
  w}_i\in\mathbb{C}^{\,5}$, as
\begin{equation}\label{eq:defA2}
{\bf M}^{(l)} ({\bf w}_1\wedge\ldots\wedge{\bf w}_l) := ({\bf M} {\bf w}_1)\wedge\ldots\wedge{\bf w}_l +
\ldots + {\bf w}_1\wedge\ldots\wedge(\bM{\bf w}_l)
\end{equation}
and it extends by linearity to the non-decomposable elements in $\bigwedge^l(\mathbb{C}^5)$. General aspects
of the numerical implementation of this theory can be found
in~\cite{ab02}. \rrbf{The general form of the matrices $\bM^{(2)}$ and
$\bM^{(3)}$ can be found in the appendix}.

To determine the three-dimensional unstable manifold for $\xi\in(-\infty,0]$, we will use~\eqref{Unk} with
$l=3$. Since the induced matrix ${\bf M}^{(3)}(\xi;E_\infty,k,s)$ inherits the differentiability and
analyticity of ${\bf M}(\xi;E_\infty,k,s)$, the following limiting matrix exists:
\[
{\bf M}_-^{(3)} (E_\infty,k,s)=
\lim_{\xi\to-\infty}{\bf M}^{(3)}(\xi;E_\infty,k,s) .
\]
The set of eigenvalues of the matrix ${\bf M}_\pm^{(3)}(E_\infty,k,s)$
consists of all possible sums of three eigenvalues of ${\bf
  M}_\pm(E_\infty,k,s)$ (see \textsc{Marcus}~\cite{marcus}).
Therefore, for $s>0$ and $k\neq 0$, there is an eigenvalue $\nu_-$ of
${\bf M}_-^{(3)}$, which is the sum of the $3$ positive eigenvalues of
${\bf M}_-$, i.e.,
\[
\nu_- = k + \frac s {v^*} -\frac{v^*}{2D} +
\frac{\sqrt{v^{*2}+4D(\sigma^-+s+Dk^2)}}{2D}
\]
(note that the subscript ``$-$'' in $\nu_-$ refers to exponentially
decaying behavior at $-\infty$, not to the sign of $\nu_-$, which is
obviously positive).  The eigenvalue $\nu_-$ is simple and has real
part strictly greater than any other eigenvalue of ${\bf M}_-^{(3)}$
(as ${\bf M}_-$ is hyperbolic).  We denote the eigenvector associated
with $\nu_-$ as ${\bf W}_e^-$, i.e.,
${\bf M}_-^{(3)}{\bf W}_e^- = \nu_-{\bf W}_e^-$.
This vector can always be constructed in an analytic way (see
\rbf{\cite[pp.\ 99-101]{kato}, \cite{derks02, bz02, humpherys06}}).
\rbf{In this case it is easy to determine an explicit analytical
  expression for the eigenvector as ${\bf M}_-$ is quite sparse.}
The unstable manifold corresponds to the solution ${\bf W}^-(\xi)$ of
the linearized system~\eqref{Unk} (with $l=3$) which satisfies
$\mijnlim{\xi\to-\infty} e^{-\nu_-\xi} {\bf W}^-(\xi) = {\bf W}_e^-$.

The stable manifold can be determined in a similar way. As indicated
in the previous section, the scaled system~\eqref{eq.linsys_scaled}
will be used to determine the stable manifold. For the stable manifold
with $\xi\in[0,\infty)$, we will use~\eqref{Unk} with $l=2$ and the
scaled matrix~$\widetilde {\bf M}$ . As before, the asymptotic matrix
\[
{\bf M}_+^{(2)}(E_\infty,k,s) =
\lim_{\xi\to\infty}\widetilde{\bf M}^{(2)}(\xi;E_\infty,k,s)\,.
\]
exists. Now the eigenvalues of ${\bf M}_+^{(2)}(E_\infty,k,s)$
consists of all possible sums of two eigenvalues of $\widetilde{\bf
  M}_\pm(E_\infty,k,s)$. Therefore, for $s>0$, $k\neq 0$, ${\bf
  M}_+^{(2)}$ has an eigenvalue~$\nu_+$, which is the sum
of the $2$ negative eigenvalues of $\widetilde {\bf M}_+$, i.e.,
\[
\nu_+= -\left(\sqrt{\frac{s + Dk^2}{D}} + k -\beta\right)
\]
As before, this eigenvalue is simple and has real part strictly less
than any other eigenvalue of ${\bf M}_+^{(2)}$. The eigenvector
associated with~$\nu_+$ will be denoted by ${\bf W}_e^+$, i.e.,
%
${\bf M}_+^{(2)}{\bf W}_e^+ = \nu_+{\bf W}_e^+$
%
The stable manifold of the scaled system~\eqref{eq.linsys_scaled}
corresponds to the solution ${\bf W}^+(\xi)$ of the linearized
system~\eqref{Unk} (with $l=2$ and $\bM=\widetilde\bM$) which
satisfies $\mijnlim{\xi\to\infty} e^{-\nu_+\xi} {\bf W}^+(\xi) = {\bf
  W}_e^+$. To get the stable manifold of the original unscaled system,
the inverse scalings matrix~${\bf D}^{-1}(\xi)$ has to be used. For
arbitrary~$\xi\geq 0$, the transformation in the wedge
space~$\bigwedge^2(\mathbb{C}^5)$ is quite complicated, but we will
only need the original stable manifold at $\xi=0$.  And at $\xi=0$,
the scalings matrix is the identity matrix. Hence at $\xi=0$, the
scaled stable manifold and the original stable manifold are the same
and ${\bf W}_e^+(0)$ describes the stable manifold
of~\eqref{eq.linsys} at $\xi=0$.

With the stable and unstable manifold as found above, the Evans function
can be defined as
\begin{equation}\label{2.9}
\Delta(E_\infty,k,s) =
{\bf W}^-(0;E_\infty,k,s)\wedge {\bf W}^+(0;E_\infty,k,s), \quad
s>0,\, k \neq 0.
\end{equation}
Thus the Evans function~$\Delta$ is more or less the determinant of
the matrix formed by a basis of the unstable manifold at $\xi=0$ and a
basis of the stable manifold at $\xi=0$. If this function is zero,
then the bases are linearly dependent, hence the two manifolds have a
non-trivial intersection.

We have focused on the case $s>0$. For $-Dk^2<s<0$, the system is
still hyperbolic, but with a two dimensional unstable manifold and a
three dimensional stable manifold. The method above can be easily
adapted to calculate the dispersion curve in this region too.

\subsection{Numerical results on the dispersion relation with the
  Evans function}\label{sec.results}

To calculate the Evans function numerically, first the front solution
has to be determined numerically as it 
\rrbf{appears explicitly in} the
linearization matrix~$\bM(\xi;E_\infty,k,s)$. The front is an
invariant manifold connecting two fixed points of the
ODE~\eqref{eq.front}, so it can be easily determined by invariant
manifold techniques or shooting, using the package
DSTool~\cite{dstool}. Shooting works in this case as the front
connects a one-dimensional unstable manifold to a three-dimensional
center-stable manifold in the ODE~\eqref{eq.front}.

After determining the fronts, the stable and unstable manifolds can be
calculated by numerical integration, \rbf{see
  e.g.~\cite{ab02,derks02,bz02}}.  In the numerical calculation of the
stable manifold, we will use $\beta=\frac12\min(\Lambda^*,k)$.  For
the stable manifold, the linearized equation on
$\bigwedge^2(\mathbb{C}^5)$
\[
\widehat{\bf W}^+_\xi = \left[\widetilde{\bf M}^{(2)}
  (\xi;E_\infty,k,s)-\nu_+(E_\infty,k,s) \bI\right]\,
\widehat{\bf W}^+ , \quad
\widehat{\bf W}^+(\xi)\big|_{\xi=L_\infty} = \bW_e^+(E_\infty,k,s)\,,
\]
is integrated from $x=L_\infty$ to $\xi=0$, using the second order
Gauss-Legendre Runge-Kutta (GLRK) method, i.e.  the implicit midpoint rule.
Here the scaling $\widehat{\bf W}^+(\xi) = e^{-\nu_+ \xi}\, {\bf
  W}^+(\xi)$ ensures that any numerical errors due to the exponential
growth are removed and $\widehat{\bf W}^+(\xi)\big|_{\xi=0}={\bf
  W}^+(\xi)\big|_{\xi=0}$ is bounded.  The
eigenvector~$\bW_e^+(E_\infty,k,s)$ can be determined explicitly as
wedge product of the relevant eigenvectors of~$\bM^+(E_\infty,s,k)$
thanks to the sparse nature of this matrix.

For the unstable manifold, the linearized equation on
$\bigwedge^3(\mathbb{C}^5)$
\[
\widehat{\bf W}^-_\xi = \left[{\bf M}^{(3)}
  (\xi;E_\infty,k,s)-\nu_-(E_\infty,s,k) \bI\right]\,
\widehat{\bf W}^- , \quad
\widehat{\bf W}^-(\xi)\big|_{\xi=L_\infty} = \bW_e^-(E_\infty,s,k)\,,
\]
is integrated from $x=-L_\infty$ to $\xi=0$, also using the implicit
midpoint rule and introducing the rescaling $\widehat{\bf W}^-(\xi) =
e^{-{\nu_-}\xi}\, {\bf W}^-(\xi)$ to remove potential exponential
growth.  Again, the eigenvector~$\bW_e^-(E_\infty,k,s)$ can be
determined explicitly as wedge product of the relevant eigenvectors
of~$\bM^-(E_\infty,s,k)$.

At $\xi=0$, the computed Evans function is (see \eqref{2.9})
\begin{equation}\label{enumeric}
\Delta(E_\infty,k,s) = {\bf W}^-(0) \wedge {\bf W}^+(0)
= \widehat{\bf W}^-(0) \wedge \widehat{\bf W}^+(0).
\end{equation}
For $s=0=k$, the center-stable and the center-unstable manifold have a
two-dimensional intersection, due to the translation and gauge
invariance, see section~\ref{sec.k_small} for details. In order to
determine the dispersion curve, we start near $k=0$ and $s=0$ and then
slowly increase $k$ and determine for which $s(k)$ the Evans function
$\Delta(E_\infty,k,s(k))$ vanishes.

The numerical errors in the calculation of the Evans function are
mainly influenced by the step size used in the numerical integration
with the GLRK method and errors in the numerically determined front.
The numerical integration uses the step size $\delta x=0.01$. We have
performed various checks with a decreased step size and this shows
that the error in the value of $s$ for fixed $k$ is largest (order
$10^{-4}$) if $k$ is small and decreases for larger $k$ (order
$10^{-6}$). The accuracy of the front has been checked and is such
that the error in the front gives a negligible error (compared to the
error due to the error in the step size) in the value of $s(k)$. It
turns out that the scheme is not very sensitive to errors in the front
(at least for the~$E_\infty$ and~$D$ values considered).

In the following sections, we will present data for the dispersion
curve for varying electric field~$E_\infty$ and diffusion
coefficient~$D$. A more detailed discussion of the data, relation with
analytical asymptotics and some empirical fitting can be found in
section~\ref{sec.6}.

\subsubsection{Varying the electric field ahead of the front}
\label{sec.ds_E_numerics}

\begin{figure}[htb]
  \subfigure[$E_\infty=-1$, $-5$ and $-10$ and fixed $D=0.1$.]
  {\includegraphics[width=0.483\textwidth]{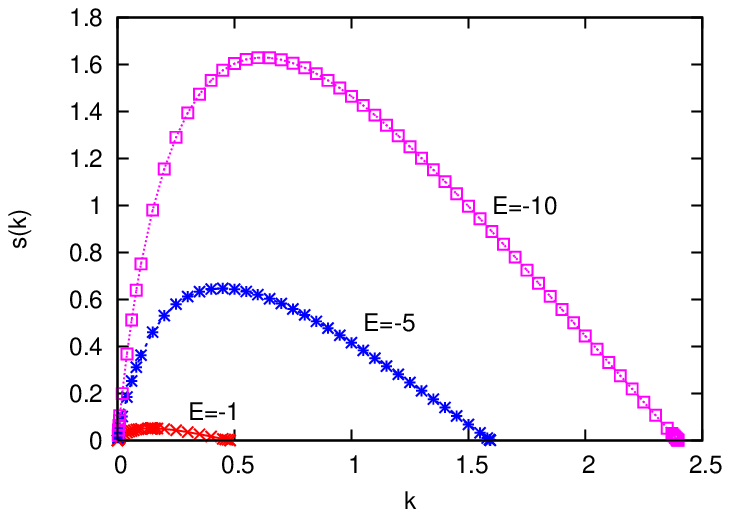}%
    \label{fig.Evarying_b}}
  \quad \subfigure[Fixed $E_\infty=-1$ and $D=0.1$, $0.01$
  and $0$.]
  {\includegraphics[width=0.483\textwidth]{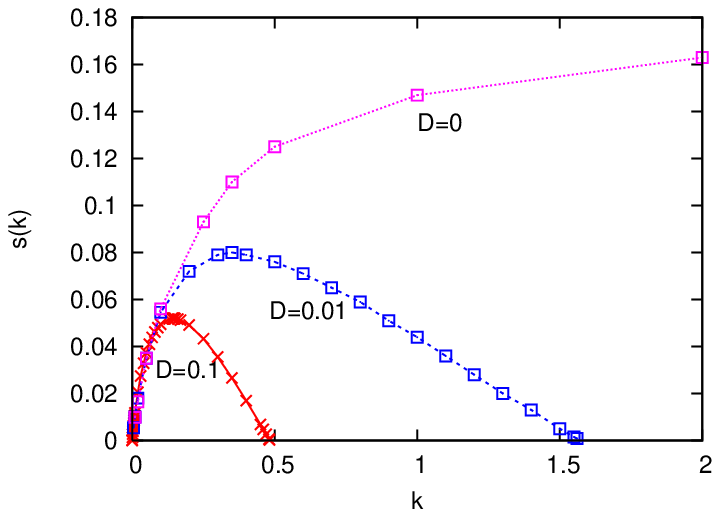}%
    \label{fig.Dvarying_b}}
  \caption{Dispersion curves $s(k)$: (a) for varying $E_\infty$ and fixed
    $D=0.1$, and (b) for fixed $E_\infty=-1$ and varying $D$. The pairs
    $(E_\infty,D)$ shown are the same as in Fig.~\ref{fig.front}. The data
    for the singular limit $D=0$ is taken from
    \cite{D=0}.\label{fig.Evarying}}
\end{figure}

First we consider how the dispersion curve depends on the electric
field $E_\infty$ ahead of the front, while the diffusion coefficient
is fixed to $D=0.1$. In Figure~\ref{fig.Evarying_b}, the dispersion
curve is shown for $E_\infty=-1$, $E_\infty=-5$ and $E_\infty=-10$.
The figure shows that the shape of the dispersion curve stays similar,
but the scales of $s$ and $k$ increase when $E_\infty$ increases. The
dispersion curves can be characterized by the maximal growth rate
$s_{\rm max}$ and the corresponding wave number $k_{\rm max}$ where
$s(k_{\rm max})=s_{\rm max}$ as well as by the wave number $k_0>0$
with $s(k_0)=0$ that limits the band $0<k<k_0$ of wave numbers with
positive growth rates.

\subsubsection{Varying the diffusion coefficient}
\label{sec.ds_D_numerics}

Next we consider the effect of varying the diffusion coefficient~$D$,
while keeping the electric field ahead of the front fixed at
$E_\infty=-1$. In~\cite{D=0} it is shown that if diffusion is ignored
($D=0$), the dispersion curve stays positive and is monotonically
increasing to the saturation value $s(k)=f(|E_\infty|)/2$ for
$k\to\infty$.  Our numerics shows that if diffusion is present, this
is not the case anymore. This is not surprising as diffusion is a
singular perturbation. In Figure~\ref{fig.Dvarying_b}, the dispersion
curve is shown for $D=0.1$, $D=0.01$ and $D=0$; the data for $D=0$ is
taken from \cite{D=0}.  It shows that the growth rate $s(k)$ has a
maximum $s_{\rm max}$ if diffusion is present and becomes negative for
$k$ larger than some $k_0$. Furthermore for decreasing diffusion $D$, the maximal growth
rate moves upward towards the saturation value $f(|E_\infty|)/2$ for $D=0$.
This suggests that some features of the dispersion curve behave
regularly in $D$, in spite of the fact that $D$ is a singular
perturbation. For example, for a finite wave number interval, the
limit of the dispersion curves for $D\to 0$ exists and is the curve
for $D=0$. However, the asymptotic profile for large values of the
wave number is obviously singular in~$D$.  This duality can also be
found in the front itself: the velocity and the profile of the
ionization density and the electric field of the uniformly translating
negative front depend regularly on $D=0$, while the profile of
the ionization density is singular, as discussed in
section~\ref{sec.mult} and shown in Fig.~\ref{fig.front}.

\section{Numerical simulation of the perturbed initial value
problem}\label{sec.simulation}

In the previous section, we have determined the dispersion relation
$s(k)$ for transversal perturbations of ionization fronts as a
temporal eigenvalue problem of the PDE system linearized about the
uniformly translating planar front.  Since we are dealing with pulled
fronts (cf.\ sections~\ref{sec.intro} and \ref{sec.mult}), the problem
is unconventional: both the velocity $v^*$ of the uniformly
translating planar front and the dispersion relation $s(k)$ of its
transversal perturbations are {unique} only if the spatial decay
constraint~(\ref{InitDecay}) is imposed. Furthermore a longitudinally
perturbed planar front approaches its asymptotic profile and velocity
algebraically slowly in time (\ref{relax}). Therefore it is worthwhile
to test the predicted dispersion relation on direct numerical
simulations of the corresponding initial value problem.

In this section, we will therefore simulate the temporal evolution of
a perturbed planar front by numerically solving the full nonlinear
PDEs~\eqref{107}-\eqref{109}, and we will determine the dispersion
curve from a number of simulations with perturbations with different
wave vectors~$k$.  This is done for far field $E_\infty=-1$ and
diffusion constant $D=0.1$.

To determine the instability curve with a simulation of the full PDE, we
parametrize the evolution
of a perturbed planar front with wave number $k$ as
\be\label{rei7} \bU(x,z,t) \approx \bU_0(\xi)+\delta \,\bU_1(\xi,t)
\,e^{ikx+st},
\quad \xi=z-v^*t,\quad\bU=(\sigma,\rho,\phi). \ee
If $\delta\,e^{st}$ is small enough,
the solution is in the linear regime, and $s(k)$ can
be determined from the evolution of the perturbation after
$\bU_1(\xi,t)$ has relaxed to some time independent function.
Therefore in the numerical simulations, we choose $\delta$ for each wave
number $k$ in such a manner that $t$ is large enough to extract
meaningful growth rates and that $\delta\,e^{st}$ is small enough that
the dynamics at the final time is still well approximated by the
linearization about the planar front.

Furthermore, an appropriate choice of the initial condition reduces the
initial transient time during
which $\bU_1(\xi,t)$ in the co-moving frame still explicitly depends on
time $t$.
Ideally, such an initial condition is of the form $\bU(x,z,0) =
\bU_0(\xi)+\delta \,\bU_1(\xi)\,\cos kx$
etc., where $\bU_1$ is a solution of the linearized system
(\ref{eq.linsys}).
To find an approximation for $\bU_1(\xi)$,
we use that the instability acts on the position of the front, i.e., we
write the perturbed front as
$\bU_0(\xi+\delta e^{ikx+st}) \approx \bU_0(\xi) + \delta
\,e^{ikx+st}\,\partial_\xi\bU_0(\xi)$.
Therefore we choose $\bU_1(\xi) = \partial_\xi\bU_0(\xi)$ and the
initial condition as
\be \label{rei3} \bU(x,z,0) = \bU_0(z)+\delta \,
\partial_z\bU_0(z)\,\cos kx. \ee
As $\partial_\xi\bU_0(\xi)$ is a solution of the linearized system for
$k=0=s$, this choice will be
very efficient for small values of~$k$ and require longer transient
times for larger $k$.

To solve the full 2D PDE, the algorithm as described
in~\cite{PRL02,Andrea} is used, while adaptive grid refinement as
introduced in \cite{Montijn} was not required. For fixed~$k$, the PDE
with initial condition~\eqref{rei3} is solved on the spatial
rectangle~$(x,z) \in [0,L_x]\times [0,L_z]$.  The length of the domain
in the transversal $x$-direction, $L_x$, is such that exactly 5 wave
lengths fit into the domain, i.e., $L_x=\frac{10\pi}{k}$, and periodic
boundary conditions are imposed in this direction by identifying $x=0$
with $x=L_x$. On the boundaries in the longitudinal $z$-direction,
Neumann conditions for the electron density are imposed. The potential
is constant far behind the front and the electric field is constant
far ahead
of the front; therefore for the potential $\phi$, the Dirichlet
condition $\phi=0$ is imposed at $z=0$, and the Neumann condition
$\partial_z\phi=-E_\infty$ at $z=L_z$ accounts for the far field ahead
of the front.

The amplitude of the perturbation is conveniently traced by the
maximum of the electron density
\be \label{smax}
\sigma_{\rm  max}(x,t)=\max_{z\in[0,L_z]}\sigma(x,z,t)
\ee
evaluated across the front. The reason is as follows.  First,
Figure~\ref{fig.front} shows the spatial profiles of planar fronts for
different electric fields $E_\infty$ and illustrates that for fixed
$D$, the maximum of the electron density $\sigma_{\rm max}$ as well as
the asymptotic density $\sigma^-$ behind the front strongly depend on
the field $E^+$ immediately ahead of the front, where for a planar
front the close and the far field are identical: $E^+=E_\infty$.
Second, the modulation of the front position leads to a modulation of
the electric field $E^+$ immediately before the front (cf.~discussion
in section~\ref{sec.phys.kleink}); therefore $\sigma_{\rm max}$ as a
function of $E^+$ is modulated as well.

\begin{figure}[htb]
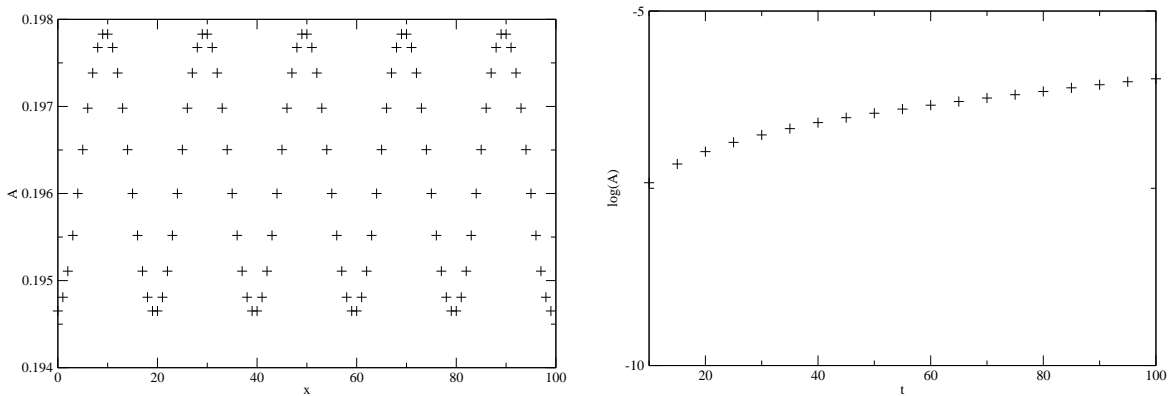

  \vspace{0.4cm} 
  \subfigure[The maximal value of the electron density $\sigma_{\rm max}
  (x,t)$ for $t=50$ as a function of the transversal coordinate
  $x$. The perturbation has wave number $k=0.45$, the transversal
  length $L_x=10\pi/k$ leaves space for 5 wave lengths that are
  clearly visible.]
  {\includegraphics[width=0.48\textwidth]{cos.eps}\label{fig.front_mod}}
  \quad \subfigure[The logarithm of the amplitude of the front
  modulation $\log A$ as a function of time~$t$ for the same $k$.]
  {\includegraphics[width=0.48\textwidth]{reimer.eps}\label{rei5}}
  \caption{Examples of data of the initial value simulation from which
    the growth rate $s(k)$ shown in Fig.~\ref{rei6} are determined.}
\end{figure}

An example of $\sigma_{\rm max}(x,t)$ as a function of the transversal
coordinate $x$ for a fixed time $t$ is plotted in
Fig.~\ref{fig.front_mod}.  The amplitude of the wave modulation is
determined by the Fourier integral
\[
A(t,k) =\frac{k}{5\pi} \int_0^{\frac{10\pi}{k}} \sigma_{\rm max}(x,t) \,
\cos kx \; dx .
\]
In Figure~\ref{rei5} we plot $\log A$ against time~$t$ for $k=0.45$.
Note that $k=0.45$ is close to $k_0=0.482$ (see
Figure~\ref{fig.Evarying_b} and Table~\ref{tab.s-k}) where the growth
rate vanishes, $s(k_0)=0$, therefore the growth rate in the present
example is small and particularly sensitive to numerical errors.

Figure~\ref{rei5} shows an initial temporal transient before steady
exponential growth is reached
(where exponential growth manifests itself as a straight line in
the logarithmic plot). This is typically observed for the larger
$k$-values ($k>0.1$); as said
before, this is related to the fact that the function $\bU_1(z)$ in the
initial condition (\ref{rei3}) is not
optimal. For $k<0.1$, there are less transients as
$\bU_1(z)\approx\partial_z\bU_0(z)$ for small values of~$k$.

\begin{figure}[htb]
\centering
\includegraphics[width=.8\textwidth]{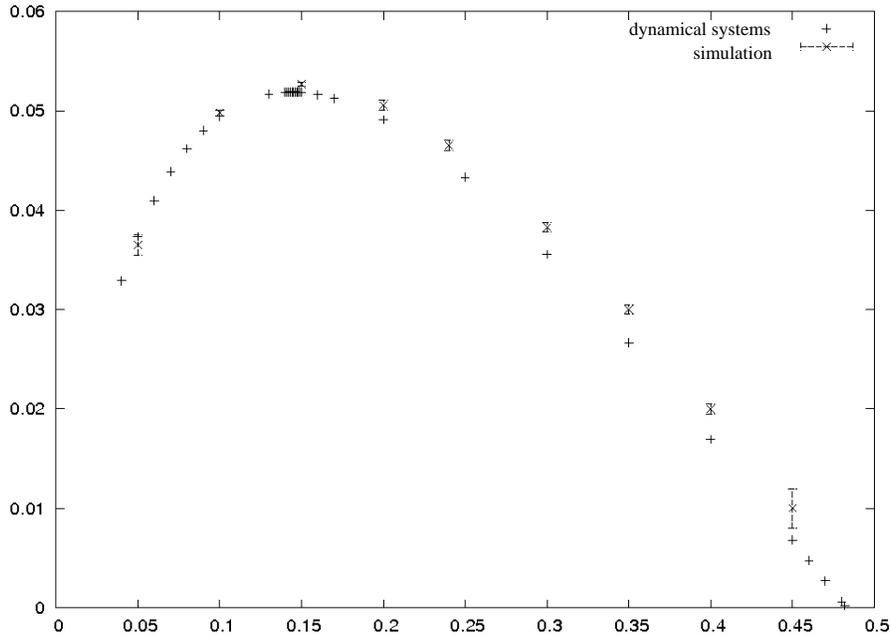}
\caption{The dispersion curve $s(k)$ for $E_\infty=-1$ and $D=0.1$. The
  crosses 
  $\times$ with error bars indicate results of simulations of
  the full initial value problem as discussed in
  section~\ref{sec.simulation} and demonstrated in
  Fig.~\ref{fig.front_mod}. For comparison, the results of the
  dynamical systems method from section~\ref{sec.results} are
  indicated with~$+$ symbols.} \label{rei6}
\end{figure}

To determine the growth rate $s(k)$, a least squares algorithm is used
to fit the best line through the
data points $(t,\log A)$, and the initial transient time
is ignored for larger values of~$k$. For each value of $k$, the growth
rate is determined with various
choices of~$\delta$. The resulting growth rate $s(k)$ is indicated in
Figure~\ref{rei6} with crosses X and
the error bars are related to the various choices of~$\delta$.

\rbf{Fig.~\ref{rei6} also shows the dispersion relation for $(E_\infty,D)=(-1,0.1)$ determined with the
dynamical systems method in the previous section~\ref{sec.results}; these numerical results are denoted with
$+$ and can now be compared with the results of the initial value problem from the present section. Around
the maximum of the curve, the agreement between the numerical results of the two very distinct methods is
convincing. For larger values of $k$, the differences increase, but the error bars of the initial value
problem results increase as well. Furthermore, the plotted error bars are an underestimation as they only
account for the errors discussed above that emerge from the choice of the initial condition and from the time
interval of evaluation and therefore from possible initial transients and from a possible transition to
nonlinear behavior. Additional errors can be due to the numerical discretization and time stepping of the
s themselves. We therefore conclude that the two results agree within the numerical error range of the
initial value simulations over the whole curve.}

\section{Analytical derivation of 
\rul{asymptotic limits} for $k\ll1$ and
  $k\gg1$}\label{asymptotics}

Having determined the dispersion relation numerically for different values of electric field~$E_\infty$ and
diffusion constant~$D$ in section~\ref{sec.3}, and having tested the correctness of the eigenvalue
calculation against numerical solutions of the initial value problem in section~\ref{sec.simulation}, we now
will analytically derive asymptotic expressions for the dispersion relation for small and large values of the
wave modes~$k$. It will be shown that these \rbf{asymptotic limits} are
\[
s(k) = \left\{
  \begin{arrayl}
     k \,E_\infty \,  \frac{dv^*}{dE_\infty}, && k\ll1\\
     -Dk^2,&&k\gg1
  \end{arrayl}\right.
\]
In doing so, we mathematically formalize and generalize the
  derivation of the small $k$ 
  \rbf{asymptotic limit} that was presented in
  \cite{D=0} for the singular limit $D=0$, and we correct the result
  proposed in \cite{ManD}; and we also rigorously derive the large $k$
  \rbf{asymptotic limit}, in agreement with the form proposed
  in~\cite{ManD}.

\subsection{Analysis for the \rul{asymptotic limit} $k\ll 1$}
\label{sec.k_small}

Translation invariance and electrostatic gauge invariance give two
explicitly known bounded solutions of the linearized
system~\eqref{eq.linsys} at $k=0$ and $s=0$. These are
\[
\bu_0' (\xi)=
(\sigma_0''(\xi),\sigma_0'(\xi),\rho_0'(\xi_),E_0'(\xi),
-E_0(\xi))^T \qmbox{and} \Be_5 = (0,0,0,0,1)^T.
\]
Note that $\Be_5$ is a solution of the linearized
system~\eqref{eq.linsys} for $k=0$ and $s$ arbitrary.

From the asymptotics of~\eqref{eq.linsys} for $k=0=s$ at
$\xi=-\infty$, we see that the only exponentially decaying solution at
$\xi=-\infty$ is given by $\bu_0'(\xi)$. This solution is related to
the only positive eigenvalue $\mu_+^-$ (see~\eqref{eq.eigenval-}). For
$\xi\to+\infty$, the solution $\bu_0' (\xi)\to -E_\infty\Be_5$, hence
this solution is not exponentially decaying for $\xi=+\infty$.
However, it is easy to obtain an explicit exponentially decaying
solution at $\xi=+\infty$, this is $\bu_0'(\xi)+E_\infty\Be_5$.

From the eigenvalues in~\eqref{eq.eigenval-} it follows that for
$0<k\ll 1$ and $0<s\ll 1$, the three-dimensional unstable manifold at
$\xi\to-\infty$ involves one eigenfunction with a fast exponential
decay (related to the eigenvalue~$\mu^-_+$) and two eigenfunctions
with a slow exponential decay (related to the eigenvalues~$k$
and~$\frac s{v^*}$).  Similarly, from the eigenvalues
in~\eqref{eq.eigenval+}, it follows that the two-dimensional stable
manifold at $\xi\to+\infty$ involves one eigenfunction with a fast
exponential decay (related to the eigenvalue~$-\Lambda^*+\beta +
\mu^+_-$) and one eigenfunction with a slow exponential decay (related
to the eigenvalue~$-k$). Recall that the stable manifold is defined as
a subset of the full stable manifold to account for the spatial decay
condition~\eqref{InitDecay}.

We focus on approximating an exponentially decaying solution on the
stable manifold. As we have seen above, in lowest order, this solution
is
\[
\bw^s(\xi;E_\infty,0,0) = \bu_0'(\xi)+E_\infty\Be_5 + \mathcal O (k+s).
\]
To determine the next order, we will use the slow behavior of the
asymptotic system and write
\[
\bw^{s}(\xi;E_\infty,k,s) = \bu_0'(\xi)+E_\infty\,
(0,0,0,k,1)\,e^{-k\xi} + kU^s_{1,k}(\xi) + sU^s_{1,s}(\xi) +
\mathcal{O}((s+k)^2).
\]
\rrbf{The second term on the right-hand side
  ($E_\infty\,(0,0,0,k,1)\,e^{-k\xi}$) is an approximation of the slow
  behavior on the stable manifold, while the other three terms are related
  to the fast decay.  Because of the slow decay, the} expansion is
only valid on a $\xi$-interval with $k\xi=o(1)$, hence $\xi$ shouldn't
be too large.

Substitution of these expressions into the linearized
system~\eqref{eq.linsys} gives that $U^s_{1,s}$ and $U^s_{1,k}$ have
to satisfy
\[
\begin{arrayl}
  \left(D_\xi-\bM(\xi;E_\infty,0,0)\right)\, U^s_{1,s} &=&
  \left(\frac{\sigma_0'}{D} , 0,
    \frac{\rho_0'}{v^*}, 0, 0\right);\\
  \left(D_\xi-\bM(\xi;E_\infty,0,0)\right)\, U^{s}_{1,k} &=&
  -E_\infty\,\left(-\frac{\tau_0-\sigma_0f'_0}{D}\,e^{-k\xi}, 0,
    \frac{\sigma_0f'_0}{v^*}\,e^{-k\xi}, 0, 1-\,e^{-k\xi}\right).
\end{arrayl}
\]
By analyzing the unperturbed system, we can find a particular solution
of the first equation.  The front solution
$\bu_0=(\tau_0,\sigma_0,\rho_0,E_0,\phi_0)$ satisfies
\[
\begin{arrayl}
\tau_0'&=&  -\frac{v^*+E_0}{D}\,\tau_0
+ (\sigma_0 - \rho_0)\frac{\sigma_0}{D}
             - \frac{\sigma_0\, f_0}{D}\\
\sigma_0' &=& \tau_0\\
\rho_0' &=& -\frac{\sigma_0 \, f_0}{v^*}\\
E_0' &=& -(\sigma_0 - \rho_0)\\
\phi_0'  &=& - E_0
\end{arrayl}
\]
Differentiating this system with respect to $E_\infty$ gives
\[
\left(D_\xi-\bM(\xi;E_\infty,0,0)\right)\,\frac{\partial
  \bu_0}{\partial E_\infty} =
\frac{dv^*}{dE_\infty}\,\left(-\frac{\tau_0}{D} , 0, \frac{\sigma_0 \,
  f_0}{(v^*)^2}, 0, 0\right)
= -\frac{dv^*}{dE_\infty}\,\left(\frac{\sigma_0'}{D} , 0,
\frac{\rho_0'}{v^*}, 0, 0\right).
\]
Hence
\[
U^s_{1,s} = -\left(\frac{dv^*}{dE_\infty}\right)^{-1}\,
\frac{\partial\bu_0}{\partial E_\infty}
\quad+\quad \mbox{a homogeneous solution.}
\]
The asymptotic behavior of $\frac{\partial\bu_0}{\partial E_\infty}$ is
\begin{equation}\label{eq.U1s}
\frac{\partial\bu_0}{\partial E_\infty} \sim (0,0,0,1,-\xi), \quad
\xi \to \infty.
\end{equation}
The polynomial growth in the $\phi_k$-component will need to be
canceled by the behavior of the other terms which involve~$k$ and
hence will give a relation between~$s$ and~$k$.

In fact, the $E_k$-components of the linearized system
(\ref{eq.linsys}) for any $s$ or $k$ can be expressed by the integral
equation (\ref{eq.E_implicit}). For all solutions on the stable
manifold, the limit $E_k(\infty)=\lim_{\xi\to\infty} E_k(\xi)$ is
well-defined, so we can write the $E_k$-components on the stable
manifold as
\[
E^s_k(\xi) =
c_2  e^{-k\xi} - \frac 12 \int_{\xi}^\infty
\left[e^{k(\xi-\eta)}+e^{-k(\xi-\eta)}\right]
\left[\rho_k(\eta)-\sigma_k(\eta)\right]\, d\eta,
\]
In the integral, $\sigma_k$ must satisfy the decay
condition~\eqref{InitDecay} on the stable manifold and hence will have
fast exponential decay. Furthermore, from~\eqref{eq.linsys}, it can be
seen that $\rho_k(\xi) = c_3 e^{\frac{s\xi}{v^*}} + \frac1{v^*}
{\displaystyle\int_{\xi_0}^\xi}
e^{-\frac{s(\xi-\eta)}{v^*}}\,\left[\sigma_0(\eta)f'_0(\eta)E_k(\eta)
  - f_0(\eta)\sigma_k(\eta)\right]\,d\eta$. As the term inside the
integral has fast exponential decay, we get that on the stable
manifold $c_3=0$ and $\rho_k$ has fast exponential decay too.  Thus
the integral in the expression for $E_k^s$ has fast exponential decay
for $\xi$ large. Since $\phi_k=-E_k'$, we get on the stable manifold
for $k$ small and $\xi$-values not too large, say $\xi \sim
k^{-\frac14}$ (hence $k\xi\sim k^{\frac34}$)
\[
\begin{arrayl}
(E^s_k,\phi^s_k)(\xi) &=& (k,1)\,e^{-k\xi} +
\rul {\mathcal{O}(e^{-\Lambda^*\xi})} \\
&=&
(k,1)\,\left
  (1-k\xi+\mathcal{O}(k\sqrt k)\right)  = (k,1-k\xi)+\mathcal{O}(k\sqrt k).
\end{arrayl}
\]

The exponentially decaying solution on the stable manifold is given by
\[
\bw^{s}(\xi;E_\infty,k,s) = \bu_0'(\xi) -
s\,\left(\frac{dv^*}{dE_\infty}\right)^{-1}\,\frac{\partial\bu_0}{\partial
  E_\infty}(\xi)+E_\infty(0,0,0,k,1)\,e^{-k\xi} + kU^s_{1,k}(\xi)
+\mathcal{O}((s+k)^2),
\]
and the arguments above show that the order~$k$ contribution in the
$(E_k,\phi_k)$-components is given fully by
$E_\infty(0,0,0,k,1)\,e^{-k\xi}$ and that $kU^s_{1,k}(\xi)$ does not
contribute to those components at this order. So it follows that the
polynomial growth in the $\phi_k$-component of
$\frac{\partial\bu_0}{\partial E_\infty}(\xi) $ as given
by~\eqref{eq.U1s} has to be canceled by the $\phi_k$-component in
$E_\infty(0,0,0,k,1)\,e^{-k\xi}$, i.e.,
$s\,\left(\frac{dv^*}{dE_\infty}\right)^{-1} = k E_\infty$ {or}
\begin{equation}
s = c^*k+{\cal O}(k^2),\qmbox{with} c^*= E_\infty \,  \frac{dv^*}{dE_\infty},
\label{small-k}
\end{equation}
and $v^*$ given in Eq.~\eqref{v*}.  Equation (\ref{small-k})
establishes the small $k$ limit of the dispersion relation~$s(k)$.

\subsection{A physical argument for the $k\ll 1$
\rul{asymptotic limit}}\label{sec.phys.kleink}

There is also a physical argument for the 
\rbf{asymptotic limit}~(\ref{small-k})
that generalizes the calculation in section IV.C of
reference~\cite{D=0} to nonvanishing $D>0$. For $k\ll 1$, the wave
length of the transversal perturbation $2\pi/k$ is the largest length
scale of the problem. It is much larger than the inner longitudinal
structure of the ionization front. On the length scale $2\pi/k$, the front
can therefore be approximated by a moving boundary between ionized and
non-ionized region at the position
\be z_f(x,t)=z_0+v^*(E_\infty)t+\delta\;e^{ikx+st}, \ee
and the local velocity of this perturbed front is
\be v(x,t)=\partial_tz_f(x,t)=v^*(E_\infty)+s\;\delta\;e^{ikx+st}.
\label{vf} \ee
The electric field in the non-ionized region is determined by
${\bf E}=-\nabla \phi$, where $\phi$ is the solution of the Laplace equation
$\nabla^2\phi=0$ together with the boundary conditions; these are
${\bf E}\to E_\infty\hat{\bf z}$ for $z\to\infty$ fixing the field far
ahead of the front and $\phi(...) = {\cal O}(k)\approx 0$ making the ionization
front almost equipotential.
(Due to gauge invariance the constant potential can be
set to zero.) The solution of this problem is
\be \phi(x,z,t)&=&-E_\infty
(z-z_0-v^*t)+E_\infty\;e^{-k(z-z_f)}\;\delta\;e^{ikx+st}+{\cal
  O}(\delta^2), \qmbox{for} z\ge z_f, \nonumber\\
E^+(x,t)&=&E_\infty+k\;E_\infty\;\delta\;e^{ikx+st}+O(\delta^2),
\qmbox{for}z=z_f,\label{E+} \ee
here $E^+(x,t)=\lim_{\epsilon\downarrow0}E(x,z_f+\epsilon,t)$ is the
electric field extrapolated onto the
boundary from the non-ionized side.

As the perturbation is linear, the front is almost planar
$\delta\ll2\pi/k$. Therefore it will propagate with
the velocity $v^*(E^+)=|E^+|+2\sqrt{Df(|E^+|)}$ (\ref{v*}) of the planar
front in the local field $E^+$.
Inserting $E^+$ from (\ref{E+}) and expanding about $E_\infty$, we get
\be v(x,t)=v^*(E^+(x,t))=v^*(E_\infty)+\partial_E v^*\Big|_{E_\infty}\;
k\;E_\infty\;\delta\;e^{ikx+st}+O(\delta^2). \label{vE} \ee
Comparison of (\ref{vE}) and (\ref{vf}) immediately gives the dispersion
relation $s=c^*k +O(k^2)$
(\ref{small-k}), that generalizes the result $s(k)=|E_\infty|k+O(k^2)$
that was derived in \cite{D=0} for the singular limit $D=0$.

\subsection{Analysis for the \rul{asymptotic limit} $k\gg
  1$}\label{sec.5.3}

The 
\rbf{asymptotic limit} for $k\gg1$ is derived by a contradiction argument. We
will suppose that $k$ is large and that $s+Dk^2$ is positive, but not
small, i.e.,
\begin{equation}\label{eq.cond_s_k}
k\gg 1, \quad s+Dk^2>0, \qmbox{and} s+Dk^2 \neq o(1)
\end{equation}
and show that this does not allow for bounded solutions. With the
assumptions above on $s$ and $k$, the dominant contributions in the
matrix ${\bf M}$ on the whole axis $\xi$ are
\be
\label{asymptM} {\bf M}_\infty =
\begin{pmatrix}
  0 & \frac{s+Dk^2}D & 0 & 0 & 0\\[1mm]
  1 & 0 & 0 & 0 & 0\\[1mm]
  0 & 0 &\frac s {v^*} & 0 & 0\\[1mm]
  0 & 0 & 0 & 0 & -k^2\\[1mm]
  0 & 0 & 0 & -1 & 0
\end{pmatrix} + O(1).
\ee
Here the three entries $-1$, $1$ and $\frac s {v^*}$ are necessary for
a nonvanishing determinant.

We want to use the Roughness Theorem~\cite{coppel78} for exponential
dichotomies to show that for~$k$ large and $s$ not close to $-Dk^2$,
the exponential dichotomy of the constant coefficient ODE is close to
the exponential dichotomy of the full system.  So first we recall the
definition of an exponential dichotomy, which gives projections on
stable or unstable manifolds.
\begin{definition}[\cite{coppel78}]
  Let $\bA$ be a matrix in $\real^{n\times n}$, $u\in\real^n$, and $J =
  \mathbb R_-$, $\mathbb R_+$, or~$\mathbb R$.  Let $\bPhi(y)$ be a
  solution matrix of the linear system
  \begin{equation}
    \label{eq.linsysM}
    \frac{du}{dy} = \bA(y) u, \quad y\in J.
  \end{equation}
  The linear system~\eqref{eq.linsysM} is said to possess an
  exponential dichotomy on the interval~$J$ if there exist a
  projection $\bf P$ and constants~$K$ and $\kappa^s<0<\kappa^u$ with the
  following properties:
\[
\begin{arrayl}
   |\bPhi(y){\bf P}\bPhi^{-1}(y_0) | &\leq&
   K\,e^{\kappa^s(y-y_0)} ,\qmbox{for} y\geq y_0, \: y,y_0\in J\\
   |\bPhi(y)({\bf I}-{\bf P})\bPhi^{-1}(y_0) | &\leq&
   Ke^{\kappa^u(y-y_0)}  ,\qmbox{for} y_0\geq  y, \: y,y_0\in J
    \end{arrayl}
    \]
\end{definition}
An extension for PDEs of this definition can be found
in~\cite{peterhof97}.

The Roughness Theorem for exponential dichotomies states the following.
\begin{theorem}[Roughness Theorem~\cite{coppel78}] \label{th.rough}
Consider the system
\begin{equation}\label{eq.as}
\frac{du}{dy} = [\bA_0 + \bA_1(y)] u,
\end{equation}
with $\bA_0\in \real^{n\times n}$ a hyperbolic matrix and $u\in\real^n$.
Then for all $\delta_0>0$ there exists a $\delta_1>0$ such that for
all matrix functions $\bA_1:\real\to\real^{n\times n} $ with
$\|\bA_1\|_{L^\infty(\real^+,\real^{n\times n})}<\delta_1$, the
system~\eqref{eq.as} has an exponential dichotomy on $\real^+$ (and
$\real^-$) with its dichotomy exponents and projections
$\delta_0$-close to those of $\frac{du}{dy} = \bA_0 u$ (in the
$L^\infty(\real^+,\real^{n\times n})$ norm).
\end{theorem}

A constant coefficient linear system does not have bounded solutions.
So if the exponential dichotomy of the linearized
system~\eqref{eq.linsys} is close to the one of the constant
coefficient system with the matrix~$\bM_\infty$ as in~\eqref{asymptM},
the linearized system~\eqref{eq.linsys} does not have bounded
solutions either. We will show that this is the case if~$s$ and~$k$
satisfy the assumptions~\eqref{eq.cond_s_k}.

First we introduce some scaling and coordinate transformations.
Define the small parameter $\eps=\frac 1 k$ and the scaled spatial
variable, the transformation matrix and transformed vector
\[
\eta = k\xi=\frac\xi\eps , \quad
{\bf T}(\eps) = \diag(\eps,1,\eps,\eps,1) \qmbox{and}
\widehat \bw(\eta) = {\bf T}(\eps)\bw(\eps\eta).
\]
Now~\eqref{eq.linsys} can be written as
\begin{equation}
\label{eq.linsys_largek}
\partial_\eta {\widehat\bw} =
\left[\widehat\bM_0(\eps,s) + \eps
  \widehat\bM_1(\eta;E_\infty,\eps) \right]\, \widehat\bw,
\end{equation}
with
\[
\widehat\bM_0(\eps,s) =
\begin{pmatrix}
  0 & 1 +\frac{\eps^2 s}{D} & 0 & 0 & 0\\
  1 & 0 & 0 & 0 & 0\\
  0 & 0 & \frac{s\eps}v & 0 & 0\\
  0 & 0 & 0 & 0 & -1\\
  0 & 0 & 0 & -1 & 0
\end{pmatrix}
\]
and
\[
\widehat\bM_1(\eta;E_\infty,\eps) =
\begin{pmatrix}
-\frac{E_0+v^*}{D} & \eps\,\frac{2\sigma_0-\rho_0-f_0}D&
-\frac{\sigma_0}D & -\frac{\partial_\xi\sigma_0 -\sigma_0 f_0'}D &
0\\[1mm]
0 & 0 & 0 & 0 & 0\\[1mm]
0 & -\eps\,\frac{f_0}{v^*} & 0 & - \frac{\sigma_0 f_0'}{v^*} & 0\\[1mm]
0 & -\eps & 1 & 0 & 0\\[1mm]
0 & 0 & 0 & 0 & 0
\end{pmatrix}, \qmbox{where} \xi=\eps\eta.
\]
The eigenvalues and eigenvectors of the constant coefficient
matrix~$\widehat\bM_0$ are
\[
\begin{arrayl}
\pm 1, &\mbox{with eigenvectors}& \bw_{\pm 1}=(0,0,0,\mp 1, 1);\\[2mm]
\pm\sqrt{1+\frac{\eps^2 s}{D}}, &\mbox{with eigenvectors}&
\bw_{\pm 2}=\left(\pm\sqrt{1+\frac{\eps^2 s}{D}},1,0,0,0\right); \\[2mm]
\frac {\eps s}{v^*}, &\mbox{with eigenvector}&
\bw_{3}=\left(0,0,1,0,0\right).
\end{arrayl}
\]
If $|s|\ll \frac 1\eps$, then the matrix $\widehat\bM_0(\eps,s)$ is
not hyperbolic at $\eps=0$. However, this problem is not fundamental
as it is known that there is a hyperbolic splitting in the full
problem (see section~\ref{sec.dyn_sys}) and there is a spectral gap
between the positive and negative eigenvalues, even if $\eps$ is close
to zero. The spectral gap disappears if $s\approx -Dk^2$ (or $\eps^2 s
\approx -D$) and in this case the following arguments will not work.
\rbf{The spectral gap allows us to define a weight function which
  moves the spectrum away from the imaginary axis.}  To be specific, define
\[
\widetilde \bw(\eta) = e^{\nu\eta} \widehat \bw(\eta) , \qmbox{with}
\nu = \left\{
\begin{arrayl}
\frac12\mathop{\rm sgn}(s) , & \mbox{if} & |\eps s|<v^*\\
0, & \mbox{if} & |\eps s|\geq v^*
\end{arrayl}
\right.
\]
Then $\widetilde \bw(\eta)$ satisfies the ODE
\begin{equation}\label{eq.linsys_largek_scaled}
\partial_\eta {\widetilde\bw} =
\left(\left[\widehat\bM_0(\eps,s)+\nu \bI \right] + \eps
  \widehat\bM_1(\eta;E_\infty,\eps) \right)\, \widetilde\bw
\end{equation}
and the spectrum of $\widehat\bM_0(\eps,s)+\nu \bI$ is bounded away
from zero for all $\eps$ small (as long as $s+Dk^2$ is not small). The
system
\begin{equation}\label{eq.linsys0}
\partial_\eta {\widetilde\bw} =
\left[\widehat\bM_0(\eps,s)+\nu \bI\right]\, \widetilde\bw
\end{equation}
has an exponential dichotomy with projection~$\bP_0$ such that the
range of $\bP_0$ is the span of all eigenvectors of the negative
eigenvalues and the kernel of $\bP_0$ is the span of all eigenvectors
of the positive eigenvalues. Then $\bP_0$ is the projection on the
stable subspace of the linear system~\eqref{eq.linsys0} and
$\bI-\bP_0$ is the projection on the unstable subspace.

Clearly the matrix $\widehat \bM_1(\eta;E_\infty,\eps)$ is uniformly
bounded for all $\eta$ and $\eps$ small.  Thus applying the Roughness
Theorem~\ref{th.rough} gives that for $\eps$ small there is an
exponential dichotomy for the system~\eqref{eq.linsys_largek_scaled}
on $\mathbb{R}^+$ with projection $\bP_\eps^s(\eta)$ onto the stable
subspace such that $\bP_\eps^s(\eta)$ is $\eps$-close to $\bP_0$ for
all $\eta \geq 0$. And similarly, there is an exponential dichotomy
for the system~\eqref{eq.linsys_largek_scaled} on $\mathbb{R}^-$ with
projection $\bP_\eps^u(\eta)$ onto the unstable subspace such that
$\bP_\eps^u(\eta)$ is $\eps$-close to $\bI-\bP_0$ for all $\eta \leq
0$. Thus the range of $\bP_\eps^s(0)$ is $\eps$-close to the range of
$\bP_0$ and the range of $\bP_\eps^u(0)$ is $\eps$-close to the range
of $\bI-\bP_0$, so the range of $\bP_\eps^s(0)$ and the range of
$\bP_\eps^u(0)$ have only a trivial intersection for~$\eps$ small.

As the weight function~$e^{\nu\eta}$ has been chosen such that no
eigenvalue crosses the imaginary axis, it affects only the value of
the dichotomy exponentials, not their sign, nor the stable and
unstable manifolds.  Hence the stable and unstable manifolds
of~\eqref{eq.linsys_largek} have a trivial intersection only. And the
same holds for the stable and unstable manifolds of~\eqref{eq.linsys},
as the only difference between the systems~\eqref{eq.linsys_largek}
and~\eqref{eq.linsys} is a scaling.  So it can be concluded that the
linear system~\eqref{eq.linsys} does not have any bounded solutions
for $\eps$ small ($k$ large) and $s$ not close to $-Dk^2$.

If $s$ is close to $-Dk^2$, i.e., $s=-\frac D{\eps^2}(1 + o(1))$, then
the matrix $\widehat A_0(\eps,s)$ has a positive and negative
eigenvalue of order $o(1)$ and the spectral gap will disappear in the
limit $\eps\to0$. So the roughness theorem can not be applied anymore
and no conclusion about bounded solutions can be drawn.

The arguments above show that only if $s=-Dk^2(1+ o(1))$, there is a
possibility for bounded solutions to exist. As the dispersion curve
indicates a bounded solution of the ODE~\eqref{eq.linsys}, this
implies that for $\eps$ near zero, hence $k$ large, the dispersion
curve is
\begin{equation}\label{large-k}
s(k) = -D k^2 \left(1 + o(1)\right), \quad k\to\infty.
\end{equation}

So far we have used that
  $s+Dk^2> 0$. If one considers the linear
  system~\eqref{eq.linsys}, an edge of the continuous spectrum is
  given by the curve $s=-Dk^2+f(|E_\infty|)$. However, one should
  include the decay condition~\eqref{InitDecay}, i.e., the
  scaling~\eqref{eq.scaling}. For any $0<\beta<\Lambda^*$ the
  edge becomes $s=-Dk^2+\beta$. By taking the limit for $\beta\to 0$,
  we see that the curve $s=-Dk^2$ is an edge of the continuous
  spectrum.  Thus with the decay condition, either the dispersion
  curve satisfies $s(k)\geq -Dk^2$ or it ends at the continuous
  spectrum.

\section{Physically guided fits to the numerical dispersion relations}\label{sec.6}

In section~\ref{sec.3} we have derived dispersion relations for a
number of fields $E_\infty$ and diffusion constants $D$ by numerically
solving an eigenvalue problem, we have confirmed these calculations by
a numerical solutions of the initial value problem in
section~\ref{sec.simulation}, and we have derived
analytical 
\rbf{asymptotic limits} to these dispersion relations in
section~\ref{asymptotics}.  This sets the stage for comparing the
numerical results to the analytical 
\rbf{asymptotic limits} and for deriving physically guided empirical
fits to the numerical dispersion curves where the
analytical 
\rbf{asymptotic limits} are not applicable.  \rrbf{ The small $k$-data
  derived in section~\ref{sec.results} and the analysis of
  section~\ref{sec.k_small} are shown to be consistent in
  section~\ref{sec.test_small}. After showing in
  section~\ref{sec.test_both} that a simple cross-over formula joining
  the asymptotic behavior of the small wave numbers with the
  asymptotic behavior of the large wave numbers is not satisfactory,
  we give a data collapse, empirical fits and arguments on relevant
  scales in section~\ref{sec.data_fits}.}

\subsection{Testing the small $k$ \rul{asymptotic limits}}\label{sec.test_small}

\begin{figure}[htb]
  \centering
  \includegraphics[width=.7\textwidth]{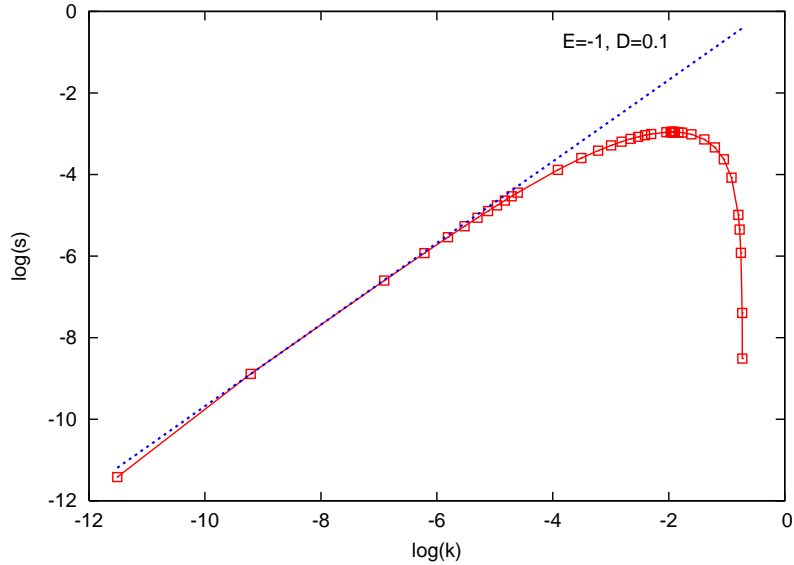}
  \caption{A $\log$-$\log$ plot of the dispersion curve $s(k)$ for
    $E=-1$ and $D=0.1$ illustrates the behavior for small~$k$. Red
    squares and solid line: Data from the numerical evaluation of the
    eigenvalue problem. Blue dashed line: Analytical 
    \rbf{asymptotic limit} $\log s = \log c^* + \log k$ according to
    (\ref{small-k}).}
  \label{fig.log_log}
\end{figure}

First the asymptotic relation (\ref{small-k}) for small $k$ is tested
on the numerical results.  Beyond the results visible in the plots of
figure~\ref{fig.Evarying}, the numerical dispersion relation for
$E_\infty=-1$ and $D=0.1$ was evaluated carefully for small values of
$k$, and the result is shown in a double logarithmic plot in
Figure~\ref{fig.log_log} that zooms in on the small $k$ behavior.
Also plotted is the analytical 
\rbf{asymptotic limit}~(\ref{small-k}).  The comparison between
numerical data and analytical 
\rbf{asymptotic limit} is convincing in the range of small $k$.

For the other values of $E_\infty$ and $D$ presented in
figure~\ref{fig.Evarying}, the dispersion relation $s(k)$ again is
fitted very well by the 
\rbf{asymptotic limit}~(\ref{small-k}) for small values of
$k$. This will be illustrated in Figure~\ref{fig.fit3} below.

\subsection{Testing both \rul{asymptotic limits}}\label{sec.test_both}

It is quite suggestive to join the small~$k$ 
\rbf{asymptotic limit}~(\ref{small-k}) with the large~$k$
\rbf{asymptotic limit}~(\ref{large-k}) into one cross-over formula
\be s(k)=c^*k-Dk^2,~~~c^*(E_\infty,D) = E\left.\frac{dv^*}{d E}\right|_{E_\infty} = |E_\infty|
\,\left(\,1+f'(|E_\infty|)\sqrt{\frac{D}{f(|E_\infty|}} \,\right). \label{cD} \ee
A formula similar to $s(k)=c^*k-Dk^2$ was suggested in \cite{ManD},
but with a different prefactor instead of $c^*$.  However, we now will
confirm once more the correctness of the prefactor $c^*$, and we will
show that the large $k$ 
\rbf{asymptotic limit} is not yet applicable in the range of positive
growth rates $s(k)$.

\begin{figure}[htb]
  \centering
  \includegraphics[width=.7\textwidth]{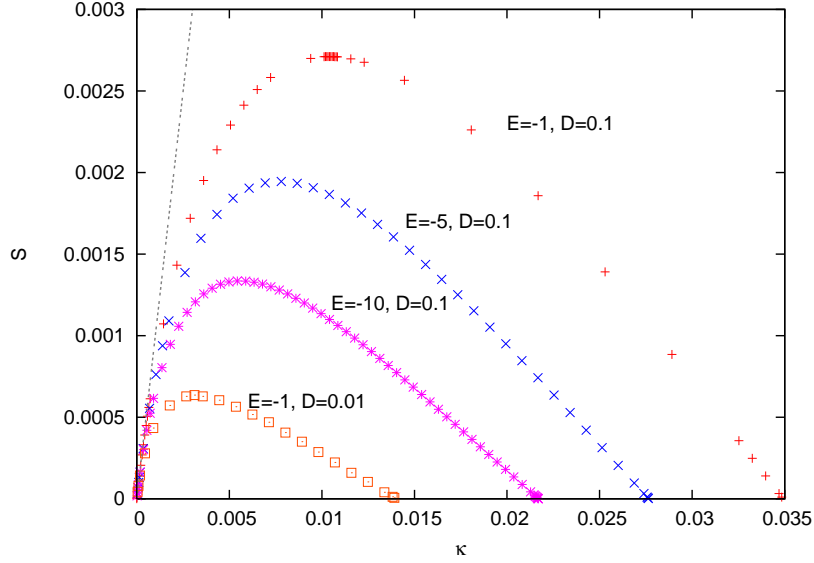}
  \caption{Labeled curves: The numerical dispersion curves from Figure~\ref{fig.Evarying}
  plotted as ${\cal S}=Ds/c^{*2}$ over $\kappa=Dk/c^*$. Dotted line on the left: the parabola $\kappa-\kappa^2$
  that would be predicted by \eqref{cD} as far as it fits into the plotted region.}
  \label{fig.fit3}
\end{figure}

If (\ref{cD}) holds, then the dispersion relation for the rescaled variable~${\cal S}=Ds/c^{*2}$ as a
function of the rescaled wave number $\kappa=Dk/c^*$ becomes ${\cal S}=\kappa-\kappa^2$.
Therefore the formula (\ref{cD}) can easily be tested on the numerical data from figure~\ref{fig.Evarying} by
plotting them in rescaled variables ${\cal S}$ and $\kappa$ with appropriate values for $D$ and
$c^*(E_\infty,D)$ for each curve.  The result is shown in Fig.~\ref{fig.fit3}, together with the parabola
$\kappa-\kappa^2$. The plot illustrates that the 
\rbf{asymptotic limit}~\eqref{small-k} indeed is a very good fit to all
data for small $k$.

For larger $k$, the curves differ quantitatively. In particular,
${\cal S}$ vanishes for $\kappa$ between 0.014 and 0.035 for the
numerical dispersion curves while the formula \eqref{cD} predicts this
to happen for $\kappa=1$. Also the maximum of the dispersion curve
${\cal S}_{\rm max}$ is never higher than 0.0027 for the numerical
data while formula (\ref{cD}) predicts 0.25. Of course, this is not in
contradiction with the analytical results in Section~\ref{sec.5.3} for
large~$k$. Rather it says that the positive part of the dispersion
curve lies completely in the range of small $k$, where the
\rbf{asymptotic limit} for large $k$ is not applicable. We conclude
that cross-over formula (\ref{cD}) is not an appropriate fit for the
numerically derived dispersion relations.

\subsection{Data collapse, relevant length scales, empirical fits and conjectures}\label{sec.data_fits}

We finish this section with a data collapse and arguments on relevant
scales that guide empirical fits.

\subsubsection{Data collapse}

\begin{figure}[htb]
  \centering
  \includegraphics[width=.7\textwidth]{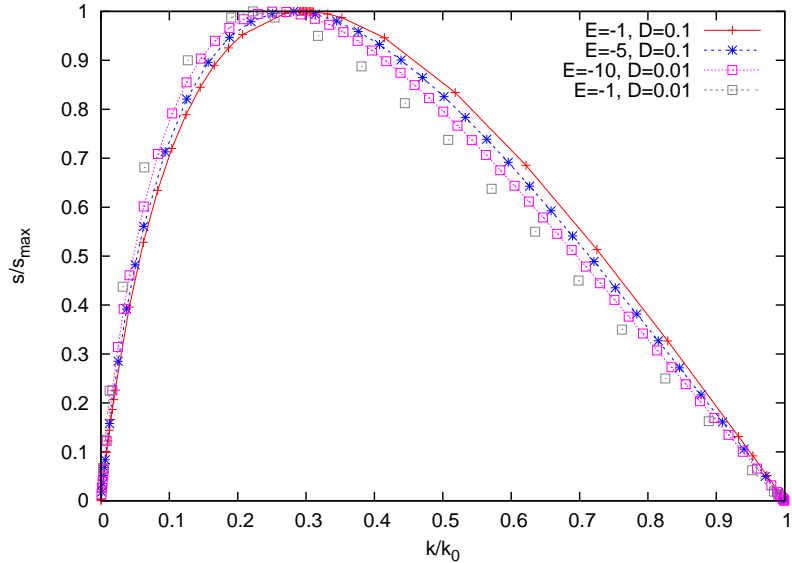}
  \caption{The numerical dispersion curves from
Figure~\ref{fig.Evarying} plotted as $s/s_{\rm max}$
  over $k/k_0$; here $s_{\rm max}=\max_k s(k)$ and $k_0>0$ with $s(k_0)=0$
are determined
  from the respective curve.}
  \label{collapse}
\end{figure}

First, we investigate \rbf{whether the numerical data for different $E_\infty$ and $D$ can be collapsed onto
one curve.} This is done by determining the maximum of the dispersion curve $s_{\rm max}$ and the wave number
$k_0$ where the growth rate vanishes, $s(k_0)=0$, from the numerical data for each pair $(E_\infty,D)$.  In
figure~\ref{collapse} all curves are plotted as $s/s_{\rm max}$ and $k/k_0$ with their respective $s_{\rm
max}$ and $k_0$. The plot shows that the curve shapes are very similar, but they do not coincide completely.
For example, there seems to be a small drift in the position of the maximum.

\subsubsection{Relevant length scales and the $D=0$ case}

In a second step, we 
\rbf{investigate} which physical or mathematical mechanisms
can suppress the growth rate $s(k)$ for much smaller values of $k$
than suggested by the large~$k$ 
\rbf{asymptotic limit}~\eqref{large-k}. In a first overview, there are
three length scales in the problem. The transversal perturbation is
characterized by its wave length $2\pi/k$. In the longitudinal
direction, the front is characterized by two length scales, the
electric screening length $\ell_\alpha$ and the diffusion length
$\ell_D$, cf.~Fig.~\ref{fig.front},
\be \ell_\alpha=\frac 1
{\alpha(E_\infty)}\quad \mbox{and}\quad
\ell_D=\frac1{\Lambda^*}=\sqrt{\frac D{f(E_\infty)}} \label{ell}.
\ee
For vanishing diffusion $D$, the diffusion length vanishes, and the
screening length $\ell_\alpha$ has to be compared to the wave length
of the perturbation; in \cite{D=0} it was shown that it determines the
cross-over from the small to the large $k$ 
\rbf{asymptotic limit} of the dispersion relation:
\be \label{asympD0} s(k)&=&\left\{\begin{array}{ll}|E_\infty|\;|k|,~~
    &\mbox{for }|k|\ll k_\alpha,\\
    |E_\infty|\;k_\alpha,~~ &\mbox{for }|k|\gg k_\alpha;\end{array}
\right. \quad\mbox{for }D=0, ~~~\mbox{where}~~~
k_\alpha=\frac1{2\ell_\alpha}.  \ee
The actual curve for $D=0$ and $E_\infty=-1$ is given in figure~\ref{fig.Dvarying_b}, where we remark that
the form
\be \label{fitD0} s(k)=\frac{|E_\infty|k}{\left[1+\left(\frac k{k_\alpha}\right)^p\right]^{1/p}} \ee
for positive real $p$ reproduces the 
\rbf{asymptotic limits}~(\ref{asympD0}), but does not fit the full
numerical curve for $E_\infty=-1$ satisfactorily for any power~$p$.
The functional form of (\ref{fitD0}) will serve below as an
inspiration for our empirical fits for $D>0$.

Searching for why \eqref{fitD0} does not properly fit the data, one realizes that there are actually two
different definitions of the screening length possible:
\be \frac 1 {\ell_\alpha^+}={\alpha(E_\infty)}\quad \mbox{and} \quad \frac 1
{\ell_\alpha^-}={\lambda^-}=\int_0^{|E_\infty|}\frac{\alpha(x)\;dx}{|E_\infty|}+{\cal O}(\sqrt D), \ee
with $\lambda^-$ from \eqref{lambda-}. The dimensions of both quantities are the same, and they approach each
other if $\alpha$ is constant in a large part of the integration interval $[0,|E_\infty|]$; this is the case
with the Townsend approximation $\alpha(x)=e^{-1/x}$ for $|E_\infty|\gg1$. Otherwise, $\ell_\alpha^+$
characterizes the slopes of the fields near the discontinuity of $\sigma$ \cite{D=0}, while $\ell_\alpha^-$
characterizes the decay \eqref{Profile-} of the fields far behind the front for $\xi\to-\infty$.
The analysis in \cite{D=0}
shows that linear perturbations with wave numbers $k\gg1$ couple to the inner local structure of the front
and are dominated by $\ell_\alpha^+$, while smaller $k$ could couple to the larger spatial structure
characterized by $\ell_\alpha^-$, this conjecture will be tested on the numerical data below and asks for
future analysis.

\subsubsection{Scales and fits for $D>0$}

When diffusion is included, the diffusion length $\ell_D$ emerges as another length scale in the front.  As
illustrated in figure~\ref{fig.front}, instead of the discontinuous electron density in the front for $D=0$,
a diffusive layer of width $\ell_D=1/\Lambda^*$ \eqref{Profile} builds up in the leading edge. While $D$
increases, the dispersion relation decreases as shown in figure~\ref{fig.Dvarying_b}. As the diffusive layer
is the main new feature of the front for $D>0$, it is plausible that the different behavior of $s$ is created
within this boundary layer. The physical mechanism is that diffusion can smear perturbations of short wave
length out, hence suppressing their growth. This process mainly takes place in the diffusive layer because
gradients are largest in this region. This idea has inspired an attempt in \cite{ManD} to calculate $s(k)$ by
local analysis within the diffusion layer. \rbf{In principle, such an approach combined with proper matched
asymptotic expansions 
could work}. However, the calculation in \cite{ManD} was intrinsically inconsistent
\cite{critique}, disagrees with our \rbf{asymptotic limit} for small~$k$ and therefore fits the numerical
results even worse than formula~(\ref{cD}), cf.~Fig.~\ref{fig.fit3}.

\begin{figure}[htb]
  \centering
  \subfigure[Using $a=3/{\alpha(|E_\infty|)}$ in formula~\eqref{eq.fit}.]%
  {\includegraphics[width=.483\textwidth]{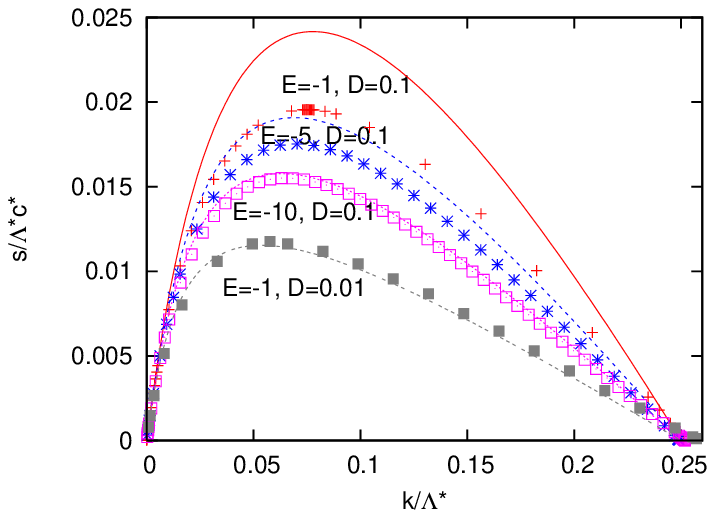}}
  \quad \subfigure[Using $a=3 c^*/{f(|E_\infty|)}$ in formula~\eqref{eq.fit}.]%
  {\includegraphics[width=.483\textwidth]{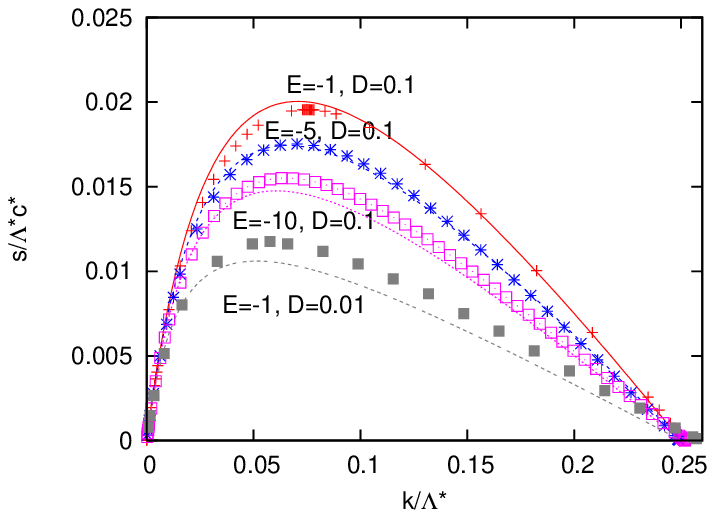}}
  \caption{The data of Figure~\ref{fig.Evarying} plotted as
    $\mathfrak{s}=s/(\Lambda^* c^*)$ over $\mathfrak{K}=k/\Lambda^*$.
The lines are given by the empirical formula~\eqref{eq.fit}.}
  \label{fig.fit5}
\end{figure}

We have tested whether the diffusion length $\ell_D=1/\Lambda^*$ plays a role in the dispersion relation by
plotting the numerical data from Fig.~\ref{fig.Evarying} this time for the rescaled variables
$\mathfrak{s}=s/(c^*\Lambda^*)$ over $\mathfrak{K} = k/\Lambda^*$. The result is shown in
Figure~\ref{fig.fit5}. It shows that the numerical dispersion curves are well approximated by
\be s(k)&\approx&c^*k+{\cal O}(k^2),\quad \mbox{for } k\to 0, \label{smallK}
\\
k_0&\approx&\Lambda^*/4,~~~\mbox{where }s(k_0)=0, \label{K0}
\ee
{The numerical evidence from Fig.~\ref{fig.fit5} summarized in~\eqref{K0} together with the physical
explanation above suggest the following conjecture.
  \begin{conjecture}\label{conj.k0}
 The largest unstable wave number of the
  Laplacian instability is proportional to the inverse diffusion length.
  \end{conjecture}
    We remark that the data gives $k_0 \approx 1/(4\ell_D)$, while the
cross-over formula (\ref{cD}) would suggest that $k_0 \approx
\ell_\alpha/\ell_D^2$, highlighting again its inadequacy for
intermediate~$k$ values.
Figure~\ref{fig.fit5} also shows that the value of the wave number for which the maximum of the dispersion
  curve is attained, lies in the range of $k_{\rm max}=0.22\;k_0$ to
  $0.30\;k_0$.}

The data in Figure~\ref{fig.fit5} suggest an empirical formula of the
form (for $s\geq0$)
\be \label{eq.fit} s(k) = \frac{c^* k}{1+a\,k} \, \left(1 - \frac{4 \,k}{\Lambda^*}\right) \qmbox{or}
\mathfrak{s} = \frac{\mathfrak{K}}{1+a\Lambda^* \mathfrak{K}}\,(1-4\mathfrak{K}) , \ee
where the parameter~$a$ will depend on the external parameters~$D$ and $E_\infty$. \rbf{The factor $c^*k$
creates the correct asymptotic limit}~(\ref{smallK}) for $k\ll1$. The factor $(1-4k/\Lambda^*)$ creates the
non-trivial zero of the dispersion relation at $k_0$ (\ref{K0}). The form of the numerator is inspired by
(\ref{fitD0}), and the proper \rbf{asymptotic limit}~(\ref{asympD0}) for large $k$ and $D=0$ would be reached
for $a=2\ell_\alpha^+ +{\cal O(\sqrt{D})}$. Obviously, the empirical formula \eqref{eq.fit} is not valid in
the asymptotic range $k\gg 1$ where $s<0$ and where the asymptotic behavior is given by $s\approx -Dk^2$.

The functional form of formula \eqref{eq.fit} is supported by the following observation. If one calculates
the maximum $(\mathfrak{K}_{\rm max},\mathfrak{s}_{\rm max})$ of~\eqref{eq.fit}, it follows that
${\mathfrak{K}_{\rm max}^2}/{\mathfrak{s}_{\rm max}} = 1/4$, independently of the value of $a$. (The number
1/4 directly stems from the factor 4 in $(1-4\mathfrak{K})$.) This relation indeed fits the numerical curves
quite well, therefore the factor 4 is supported twice independently. Relevant numerical data for this and
other fits is collected in Table~\ref{tab.s-k}.

The value for $a$ is less obvious. The empirical formula~\eqref{eq.fit} gives the following relation
between~$a\Lambda^*$ and the maximum of curve
\[
\frac{1-8\mathfrak{K}_{\rm max}}{4\mathfrak{K}^2_{\rm max}} = a\Lambda^* = \frac{1-4\sqrt{\mathfrak{s}_{\rm
max}}}{\mathfrak{s}_{\rm max}}.
\]
The empirical values for those quotients are given in Table~\ref{tab.s-k}.

The limit of $D=0$ and $k\gg1$ suggests $a=2\ell_\alpha^+ +{\cal O(\sqrt{D})}$, but the fit is unconvincing
(cf. also the discussion for $D=0$ above).  However, we found that $a = 3\ell_\alpha^+$ fits the data
reasonably well. Formula \eqref{eq.fit} with this value of $a$ together with the numerical data are
visualized in Figure~\ref{fig.fit5}(a). The fit is quite good for the lower two curves, the upper two display
some discrepancies. The main problem is the value of $\mathfrak{s}_{\rm max}$ with a relative error between
$2\%$ and $24\%$, while the position of $\mathfrak{K}_{\rm max}$ has a relative error as low as $0.5\%$ to
$7\%$. As $f(|E_\infty|)/{c^*} = \alpha(|E_\infty|) + \mathcal{O}(\sqrt{D})$, another possible fit is
$a=3c^*/{f(|E_\infty|)}$, it is displayed in Figure~\ref{fig.fit5}(b). The fit is quite good for the upper
two curves, but now the fit has some discrepancies for the lower two curves. And the value of
$\mathfrak{K}_{\rm max}$ has a larger error between $2\%$ and $10\%$ while the position of $\mathfrak{s}_{\rm
max}$ has a much smaller error of only $1\%$ to $10\%$. Obviously, these observations ask for further
analytical investigation. Note finally the striking relation between $\lambda^-=1/\ell_\alpha^-$ and the
value of $k_{\rm max}$ for larger values of the electric field in Table~\ref{tab.s-k}. As a basis for future
work, all characteristic numerical data is collected in this table.

\begin{table}[htb]\caption{\label{tab.s-k} Upper block: characteristic
numerical data of the dispersion
    relations in Figure~\ref{fig.Evarying} with errors. Middle block:
characteristic scales of the planar
  front according to analysis. Lower block: relevant ratios of numerical
and analytical scales as used for the derivation of the empirical formula.}
  \centering\renewcommand{\arraystretch}{1.3}
\begin{tabular}{c||c|c|c|c|c|}
    $(E_\infty,~D)$ & $(-1,~0.01)$ & $(-1,~0.1)$ & $(-2,~0.1)$ &
$(-5,~0.1)$ & $(-10,~0.1)$\\\hline\hline
    $s_{\rm max}$ & 0.080(1) & 0.05190(2) & 0.1695(15) & 0.647(1) &
1.6305(15) \\\hline
    $k_{\rm max}$ & 0.35(4) & 0.144(1) & 0.25(4) & 0.45(4) & 0.60(4)
\\\hline
    $k_0$ & 1.575(15) & 0.4825(5) & 0.875(25) & 1.595(1) & 2.397(1) \\\hline
    $k_0/k_{\rm max}$ & 4.56(56) & 3.35(3) & 3.60(65) & 3.57(33) & 4.01(27) \\\hline\hline
    $v^*$ & 1.12 & 1.38 & 2.70 & 6.28 & 11.9 \\\hline
    $c^*$ & 1.12 & 1.38 & 2.52 & 5.77 & 11.0 \\\hline
    $\Lambda^*=\sqrt{f(|E_\infty|)/D}$ & 6.07 & 1.92 & 3.48 & 6.40 &
    9.51 \\\hline
$\alpha(|E_\infty|)=1/\ell_\alpha^+$ &0.37 & 0.37 &0.61 & 0.82 & 0.90\\\hline
$\sigma^-$ &  0.148 & 0.144 & 0.638 &2.832 & 7.169\\\hline
$\lambda^-=1/\ell_\alpha^-$ 
& 0.13 & 0.10 & 0.23 & 0.45 & 0.60\\\hline
$3\Lambda^*/\alpha(|E_\infty|)$ & 49.5& 15.6& 17.2& 23.4&31.5\\\hline
$3\Lambda^*c^*/f(|E_\infty|)$ & 55.5& 21.6& 21.7&27.0 &34.8\\\hline \hline
$\mathfrak{K}_0 = k_0/\Lambda^*$ & 0.260(3) & 0.252(1) & 0.251(5) & 0.249(1) &
0.252(1) \\\hline
$\mathfrak{s}_{\rm max}=s_{\rm max}/(c^*\Lambda^*)$
    & 0.0118(2) & 0.0196(1) & 0.0193(2) & 0.0175(1) & 0.0155(1)
\\\hline
$k_{\rm max}/k_\alpha$ & 1.9(2) & 0.78(1) & 0.8(2) & 1.1(1) & 1.3(1) \\\hline
$\mathfrak{K}_{\rm
max}^2/\mathfrak{s}_{\rm max} $
& 0.28(7) & 0.288(4) & 0.27(9)&0.28(5) &0.26(4)
\\\hline
$(1-8\mathfrak{K}_{\rm max})/(4\mathfrak{K}_{\rm max}^2)$ & 40(6) & 17.7(1)&
21(2)& 22(2)& 31(2)
\\\hline
$(1-4\sqrt{\mathfrak{s}_{\rm max}})/\mathfrak{s}_{\rm max}$ &48.1(4) &22.5(1) &23.0(1) &26.8(1) &
32.3(1)\\\hline  \hline
  \end{tabular}
\end{table}

\section{Conclusion and outlook}

In this paper, we have found dispersion curves for negative streamer
ionization fronts by numerically solving an eigenvalue problem, we
have verified this prediction on the numerical solution of an initial
value problem, we have derived analytical expressions for the
asymptotics of the curve for large and small wave lengths, and we have
presented a physically motivated fit formula to the numerical curves
for intermediate wave lengths. The investigation is of interest for
two reasons: because pulled fronts like these ones are mathematically
challenging to investigate, and because explicit predictions on the
linear \rbf{stability}
of ionization fronts help to interpret numerical
and experimental observations of propagating and branching streamer
discharges.

The ionization front is a pulled front, i.e., the front is part of a
family of traveling waves, which
propagate into a temporally unstable steady state. For the dynamics with
one spatial variable, most traveling
waves in this family are attractors only for waves with exactly the same
asymptotic decay profile. The
exception is the pulled front, which has the steepest decay of all waves
in the family and is an attractor
for waves with a sufficiently fast decay (therefore excluding the slower
decay rates for the other traveling
waves in the family).  The instability of the state ahead of the front
and the related spatial decay
condition imply that only a submanifold of the stable manifold in the
transverse instability problem is
relevant for the transverse instability analysis. This submanifold is
identified by introducing a weighted
solution space that excludes solutions with a too slow decay rate. We
have integrated the relevant stable
submanifold and unstable manifold numerically with a dynamical systems
method to calculate the dispersion
curve. This method of finding the dispersion curve does not use any
details of the streamer model, except
that it has a pulled front. The definition of a submanifold of the
stable manifold and the subsequent
numerical integration of this stable submanifold and the unstable
manifold are ideas that can be applied to
pulled fronts in other systems, too.

{It is interesting to see that the band of unstable wave numbers
seems to be limited by a multiple of the decay rate $\Lambda^*$
that characterizes the leading edge of the pulled ionization front;
though the evidence up to now
is only numerical.} As such behavior is physically reasonable,
{the next step would  be}
to derive it analytically,
e.g., by a local analysis in the diffusive layer and matched asymptotic
expansions. Such an expansion could
be based on the limiting case where the diffusion length
$\ell_D=1/\Lambda^*$ is much smaller than the
screening length $\ell_\alpha$.

The calculated dispersion curves also contribute to understanding the
\rbf{stability}
of actual streamers. Two- and three-dimensional time
dependent simulations \cite{PRL02,Andrea,Montijn1,Montijn,Luque} of
the streamer model introduced in section 2 show them to become
unstable and branch.  Can the unstable wave lengths of this branching
be related to the unstable band of wave lengths of the present
calculation? Furthermore, if the inner front structure is approximated
by a moving boundary \cite{Bernard2,arxiv}, how is the calculated
dispersion relation of transversal perturbations to be taken into
account?  {It will also be interesting to see whether the dispersion
  relation calculated for the present fluid model is also applicable
  to the corresponding particle model~\cite{Chao}.}

Finally, we mention that the extension of the streamer model with
photo-ionization as an additional reaction
term \cite{Luque} in composed gases like air requires an extension of
the present analysis {as nonlocal interaction terms play a role}.

\appendix
\section{\rrul{Matrices in exterior algebra spaces}}\label{appendix}

\rrbf{In this appendix, we give explicit expressions for the matrices~${\bf
  M}^{(l)}$ acting on the exterior algebra
space~$\bigwedge^l(\mathbb{C}^5)$ for $l=2,3$. Let $\Be_1,\ldots,\Be_5$ be the
standard basis for $\mathbb{C}^5$. Then an induced basis on
~$\bigwedge^2(\mathbb{C}^5)$ is given by }
\begin{eqnarray*}
{\bf a}_1 &=& {\bf e}_1\wedge{\bf e}_2 \,,\quad
{\bf a}_2 = {\bf e}_1\wedge{\bf e}_3 \,,\quad
{\bf a}_3 = {\bf e}_1\wedge{\bf e}_4 \,,\quad
{\bf a}_4 = {\bf e}_1\wedge{\bf e}_5 \,,\quad
{\bf a}_5 = {\bf e}_2\wedge{\bf e}_3 \,,\nonumber\\
{\bf a}_6 &=& {\bf e}_2\wedge{\bf e}_4 \,,\quad
{\bf a}_7 = {\bf e}_2\wedge{\bf e}_5 \,,\quad
{\bf a}_8 = {\bf e}_3\wedge{\bf e}_4 \,,\quad
{\bf a}_9 = {\bf e}_3\wedge{\bf e}_5 \,,\quad
{\bf a}_{10} = {\bf e}_4\wedge{\bf e}_5 \,.
\end{eqnarray*}

\rrbf{The matrix ${\bf M}^{(2)}:\bigwedge^2(\mathbb{C}^5) \to
\bigwedge^2(\mathbb{C}^5)$ can be associated 
with a complex $10\times 10$ matrix with entries such that 
\begin{equation}\label{entries1}
{\bf M}^{(2)} \ba_i = \sum_{j=1}^{10} {\bf M}^{(2)}_{ij} \ba_j
,\quad i,j=1,\ldots,10\,, 
\end{equation}
where, for any decomposable
${\bf x}={\bf x}_1\wedge{\bf x}_2\in \bigwedge^2(\mathbb{C}^5)$,
${\bf M}^{(2)}{\bf x} := {\bf M}{\bf x}_1\wedge{\bf x}_2 +
{\bf x}_1\wedge{\bf M}{\bf x}_2$.
Let ${\bf M}$ be an arbitrary $5\times 5$ matrix with complex
entries,
\begin{equation}\label{Mdef}
{\bf M} = \begin{pmatrix}
m_{11} & m_{12} & m_{13} & m_{14} & m_{15} \\
m_{21} & m_{22} & m_{23} & m_{24} & m_{25} \\
m_{31} & m_{32} & m_{33} & m_{34} & m_{35} \\
m_{41} & m_{42} & m_{43} & m_{44} & m_{45} \\
m_{51} & m_{52} & m_{53} & m_{54} & m_{55}
\end{pmatrix}\,,
\end{equation}
then ${\bf M}^{(2)}$ takes the explicit form
\[
{\bf M}^{(2)} = \left[\begin{matrix}
d_{12}& m_{23} & m_{24} & m_{25} & -m_{13} &
 -m_{14} & -m_{15} & 0 & 0 & 0\\
m_{32} & d_{13} & m_{34} & m_{35} & m_{12} &
0 & 0 & -m_{14} & -m_{15} & 0\\
m_{42} & m_{43} & d_{14} & m_{45} & 0 &
 m_{12} & 0 & m_{13} & 0 & -m_{15}\\
m_{52} & m_{53} & m_{54} & d_{15} & 0 &
0 & m_{12} & 0 & m_{13} & m_{14}\\
-m_{31}& m_{21}& 0 & 0 & d_{23} &
m_{34} & m_{35} & -m_{24} & -m_{25} & 0\\
-m_{41}& 0 & m_{21}& 0 & m_{43} &
d_{24} & m_{45} & m_{23} & 0& -m_{25}\\
-m_{51}& 0 & 0 & m_{21}& m_{53} &
m_{54} & d_{25} & 0 & m_{23} & m_{24}\\
0      & -m_{41}& m_{31}& 0 & -m_{42} &
m_{32} & 0 & d_{34}& m_{45} & -m_{35}\\
0      & -m_{51}& 0 & m_{31}& -m_{52} &
0 & m_{32} & m_{54} & d_{35}  & m_{34}\\
0      & 0 & -m_{51}& m_{41}& 0 &
-m_{52} & m_{42} & -m_{53} & m_{43} & d_{45}
\end{matrix}\right]
\]
where $d_{ij} = m_{ii}\!+\! m_{jj} $.}

\rrbf{
In a similar way, the matrix ${\bf M}^{(3)}:\bigwedge^3(\mathbb{C}^5) \to
\bigwedge^3(\mathbb{C}^5)$ can be associated 
with a complex $10\times 10$ matrix. First we define an induced basis
on~$\bigwedge^3(\mathbb{C}^5)$ by
\begin{eqnarray*}
{\bf b}_1 &=& {\bf e}_1\wedge{\bf e}_2\wedge{\bf e}_3  \,,\quad
{\bf b}_2 = {\bf e}_1\wedge{\bf e}_2\wedge{\bf e}_4 \,,\quad
{\bf b}_3 = {\bf e}_1\wedge{\bf e}_2\wedge{\bf e}_5 \,,\quad
{\bf b}_4 = {\bf e}_1\wedge{\bf e}_3\wedge{\bf e}_4 \,,\nonumber\\
{\bf b}_5 &=& {\bf e}_1\wedge{\bf e}_3\wedge{\bf e}_5  \,,\quad
{\bf b}_6 = {\bf e}_1\wedge{\bf e}_4 \wedge{\bf e}_5\,,\quad
{\bf b}_7 = {\bf e}_2\wedge{\bf e}_3\wedge{\bf e}_4 \,,\quad
{\bf b}_8 = {\bf e}_2\wedge{\bf e}_3\wedge{\bf e}_5 \,,\nonumber\\
{\bf b}_9 &=& {\bf e}_2\wedge{\bf e}_4\wedge{\bf e}_5  \,,\quad
{\bf b}_{10} = {\bf e}_3\wedge{\bf e}_4\wedge{\bf e}_5 \,.
\end{eqnarray*}
The matrix for ${\bf M}^{(3)}\in\mathbb{C}^{10\times10}$ has entries
such that 
\begin{equation}\label{entries2}
{\bf M}^{(3)} {\bf b}_i = \sum_{j=1}^{10} {\bf M}^{(3)}_{ij} {\bf b}_j
,\quad i,j=1,\ldots,10\,, 
\end{equation}
where, for any decomposable
${\bf x}={\bf x}_1\wedge{\bf x}_2\wedge{\bf x}_3\in \bigwedge^3(\mathbb{C}^5)$,
${\bf M}^{(3)}{\bf x} := {\bf M}{\bf x}_1\wedge{\bf x}_2\wedge{\bf x}_3 +
{\bf x}_1\wedge{\bf M}{\bf x}_2\wedge{\bf x}_3 +{\bf x}_1\wedge{\bf
  x}_2\wedge{\bf M}{\bf x}_3 $. 
If $M$ is given by~\eqref{Mdef}, then ${\bf M}^{(3)}$ takes the
explicit form 
\[{\bf M}^{(3)} = 
\left[\begin{matrix}
d_{123} & m_{43} & m_{53} & -m_{42} & -m_{52} &
 0 & m_{41} & m_{51} & 0 & 0\\
m_{34} & d_{124} & m_{54} & m_{32} &0  &
-m_{52} & -m_{31} & 0 & m_{51} & 0\\
m_{35} & m_{45} & d_{125} & 0 & m_{32} &
 m_{42} & 0 & -m_{31} &-m_{41} &  0\\
-m_{24} & m_{23} &0  & d_{134} & m_{54} &
-m_{53} & m_{21}& 0 & 0  & m_{51}\\
-m_{25}&  0 &m_{23}& m_{45} & d_{135}  & 
m_{43} &0& m_{21}  & 0& -m_{41}\\
0 & -m_{25}& m_{24}& -m_{35} & m_{34} &d_{145}
& 0& 0& m_{21} & m_{31}\\
m_{14}&  -m_{13}&0 &  m_{12} &0 & 0 &
 d_{234} & m_{54} & -m_{53}& m_{52}\\
m_{15}& 0 & -m_{13}& 0 & m_{12} &
0 & m_{45} & d_{235} & m_{43} & -m_{42}\\
0  & m_{15}& -m_{14}& 0 & 0 & m_{12} &-m_{35} & 
m_{34} & d_{245}   & m_{32}\\
0      & 0 & 0 &m_{15}& -m_{14}& 
m_{13} & m_{25} & -m_{24} & m_{23} &
d_{345}
\end{matrix}\right]
\]
where $d_{ijk} = m_{ii}\!+\! m_{jj}\!+\! m_{kk} $.}

\section*{Acknowledgments}
We thank Bj\"orn Sandstede for helpful discussions on the roughness theorem.
We thank Willem Hundsdorfer and Ren\'e Reimer for advice and help
with the simulations in section~\ref{sec.simulation}.

\end{document}